\documentclass[prd,aps,showpacs,floats]{revtex4}
\usepackage{amsfonts}
\usepackage{amsmath}
\usepackage{amssymb}

\input{tcilatex}
\begin{document}

\title{ON THE POINCARE GAUGE THEORY OF GRAVITATION}
\author{S. A. Ali$^{1}$, C. Cafaro$^{2}$, S. Capozziello$^{3}$, Ch. Corda$%
^{4}$}
\affiliation{$^{1}$Department of Physics, State University of New York at Albany, 1400
Washington Avenue, Albany, NY 12222, USA\\
$^{2}$Dipartimento di Fisica, Universit\`{a} di Camerino, I-62032 Camerino,
Italy\\
$^{3}$Dipartimento di Scienze Fisiche, Universit\`{a} di Napoli
{}\textquotedblleft Federico II\textquotedblright\ and INFN Sez. di Napoli,
Compl. Univ. di Monte S. Angelo, Edificio G, Via Cinthia, I-80126, Napoli,
Italy\\
$^{4}$Associazione Scientifica Galileo Galilei,Via Pier Cironi 16, I-59100
Prato, Italy}

\begin{abstract}
We present a compact, self-contained\ review of the conventional gauge
theoretical approach to gravitation based on the local Poincar\'{e} group of
symmetry transformations. The covariant field equations, Bianchi identities
and conservation laws for angular momentum and energy-momentum are obtained.
\end{abstract}

\maketitle

\section{Introduction}

From the viewpoint of Classical physics, our spacetime is a four-dimensional
differential manifold. In special relativity, this manifold is Minkwoskian
spacetime $M_{4}$. In general relativity, the underlying spacetime is curved
so as to describe the effect of gravitation. Utiyama (1956) \cite{Utiyama}
was the first to propose that general relativity can be seen as a gauge
theory based on the local Lorentz group $SO(3$, $1)$ in much the same manner
Yang-Mills gauge theory (1954) \cite{Yang-Mills} was developed on the basis
of the internal isospin gauge group $SU(2)$. In this formulation, the
Riemannian connection is obtained as the gravitational counterpart of the
Yang-Mills gauge fields. While $SU(2)$ in the Yang-Mills theory is an
internal symmetry group, the Lorentz symmetry represents the local nature of
spacetime rather than internal degrees of freedom. The equivalence principle
asserted by Einstein for general relativity requires local spacetime
structure be identified with Minkowski space possessing Lorentz symmetry. In
order to relate local Lorentz symmetry to curved spacetime, we need to
solder the local (tangent) space to the external (curved) space. The
soldering tools are the so-called tetrad fields. Utiyama regarded the
tetrads as objects given \textit{a priori}. Soon after, Sciama (1962) \cite%
{Sciama} recognized that spacetime should be endowed with torsion in order
to accommodate spinor fields. In other words, the gravitational interaction
of spinning particles required a modification of the Riemannian geometry of
general relativity to be non-Riemannian; that is, curved space with torsion.
Although Sciama used the tetrad formalism for his gauge-like handling of
gravitation, his theory fell short in treating tetrad fields as gauge
fields. Kibble (1961) \cite{Kibble} made a comprehensive extension of
Utiyama's gauge theory of gravitation by showing that local Poincar\'{e}
symmetry $SO(3$, $1)\rtimes T\left( 3\text{, }1\right) $ ($\rtimes $
represents the semi-direct product) can generate a space with torsion as
well as curvature. The gauge fields introduced by Kibble's scheme include
the tetrads as well as the local affine connection. There has been a variety
of gauge theories of gravitation based on different local symmetry groups 
\cite{Grignani, Hehl, Inomata, Ivanov1, Ivanov2, Mansouri1, Mansouri2,
Sardanashvily, Chang, Neeman, Toller, Cognola}. In this review, mainly
following Kibble's approach, we demonstrate how gravitation can be
formulated from the gauge theoretical point of view.

The article is organized as follows. In Section 2, the Euler-Lagrange
equations are obtained by requiring invariance of the action integral under
variation of coordinates. In Section 3, the conservation laws of
energy-momentum and angular momentum are obtained from the vanishing
variation of the Lagrangian density under global Poincar\'{e}
transformations. Invariance of the Lagrangian density under local Poincar%
\'{e} transformations is considered in Section 4, where it is found that
invariance is preserved provided one introduces gauge fields with components 
$e_{\mu }^{\text{ }i}$ and $\Gamma _{\alpha \beta }^{\gamma }$, where the
former is interpreted as tetrads (which set the local coordinate frame) and
the latter as local affine connections defined with respect to the tetrad
frame. In Section 5, with the aid of tetrads and local affine connections,
the scheme for manipulating vectors and spinor valued fields is developed.
The explicit form of the curvature and torsion of the underlying spacetime
manifold is obtained in Section 6. In Section 7, the equation of motion for
the spinor as well as the field equations for gravity is derived using a
standard variational calculus. Our conclusions are presented in Section 8.

Before proceeding to the main discussion, we introduce the notation to be
used throughout the article. The metric in Minkowskian spacetime $M_{4}$ is
denoted by $\eta _{ij}=e_{i}\cdot e_{j}~(i$, $j=0$, $1$, $2$, $3)~$ with $%
\eta _{00}=-\eta _{11}=-\eta _{22}=-\eta _{33}=1$ and $\eta _{ij}=0$ for $%
i\neq j$. The orthonormal (Lorentz) basis vectors\textbf{\ }$e_{i}$\textbf{\ 
}are defined by\textbf{\ }$e_{i}:=\partial _{i}=\frac{\partial }{\partial
x^{i}}$. The metric of curved spacetime is given by $g_{\mu \nu }=e_{\mu
}\cdot e_{\nu }~(\mu $, $\nu =0$, $1$, $2$, $3)$, where\textbf{\ }$e_{\alpha
}=e_{\alpha }^{\;i}\left( x\right) e_{i}$.\textbf{\ }The quantities $%
e_{\sigma }^{\;i}\left( x\right) $ are called tetrads. The tetrads are
coefficients of the dual ($1$-form) basis non-holonomic co-vectors $%
\vartheta ^{a}\left( x\right) =e_{\text{ }\gamma }^{a}(x)dx^{\gamma }$. The
inverse of $e_{i}\,^{\mu }$ is denoted by $e^{i}\,_{\mu }$ and satisfies the
following orthogonality relations 
\begin{equation}
e^{i}\,_{\mu }e_{i}\,^{\nu }=\delta _{\mu }\,^{\nu }\text{,}~~~e^{i}\,_{\mu
}e_{j}\,^{\mu }=\delta ^{i}\,_{j}\text{.}
\end{equation}%
The tetrads constitute transformation matrices that map from local Lorentz
(with non-holonomic coordinates $x^{a}$) to world (with holonomic
coordinates $x^{\mu }$) bases, i.e., $v^{\alpha }=e_{i}^{\alpha }v^{i}$ with 
$v^{i}=v^{\alpha }e_{\alpha }^{i}$. The components $e_{\alpha }^{\;i}\left(
x\right) $ and $e_{i}^{\;\alpha }\left( x\right) $ transform as covariant
and contravariant vectors (under the Poincar\'{e} group) of the frame $%
x^{\mu }$, if and only if the rotations $\partial _{\lbrack \mu }e_{\lambda
]}^{\;a}$\ vanish at all points (the square brackets denote
anti-symmetrization). The equations $\partial _{\lbrack \mu }e_{\lambda
]}^{\;\ \ i}=0$\ are the so-called integrability conditions \cite{Kibble}.
If the integrability conditions are satisfied, then the tetrad takes the
form $e_{\mu }^{\;i}\left( x\right) =\partial x^{i}/\partial x^{\mu }$. The
metrics $\eta _{ij}$ and $g_{\alpha \beta }$ are related via 
\begin{equation}
g_{\mu \nu }=e_{\mu }\cdot e_{\nu }=e_{\mu }^{i}\left( x\right) e_{i}\cdot
e_{\nu }^{j}\left( x\right) e_{j}=e_{\mu }^{i}\left( x\right) e_{\nu
}^{j}\left( x\right) e_{i}\cdot e_{j}=e_{\mu }^{i}\left( x\right) e_{\nu
}^{j}\left( x\right) \eta _{ij}\text{.}  \label{metric}
\end{equation}


\section{Invariance Principle}

As is well-known, field equations and conservation laws of a given theory
can be obtained from the principle of least action. The same principle is
the basis of gauge theories. Thus, we begin with the principle of the least
action and Noether's theorem. Let $\chi (x)$ be a multiplet field defined at
a spacetime point $x$ and let $\mathcal{L}\{\chi (x)$, $\partial _{j}\chi
(x) $; $x\}$ be the Lagrangian density of the system. The action integral $I$
of the system over a spacetime $4$-volume $\Omega $ is defined by 
\begin{equation}
I(\Omega )=\int_{_{\Omega }}\mathcal{L}\{\chi (x)\text{, }\partial _{j}\chi
(x)\text{; }x\}\,d^{4}x\text{.}
\end{equation}%
Now we consider the infinitesimal variations of the coordinates 
\begin{equation}
x^{i}\rightarrow x^{\prime }{}^{i}=x^{i}+\delta x^{i}\text{,}
\end{equation}%
and the field variables%
\begin{equation}
\chi (x)\rightarrow \chi ^{\prime }(x^{\prime })=\chi (x)+\delta \chi (x)%
\text{.}
\end{equation}%
Correspondingly, the variation of the action integral is given by 
\begin{equation}
\delta I=\int_{_{\Omega ^{\prime }}}\mathcal{L}^{\prime }(x^{\prime
})\,d^{4}x^{\prime }-\int_{_{\Omega }}\mathcal{L}(x)\,d^{4}x=\int_{_{\Omega
}}\left[ \mathcal{L}^{\prime }(x^{\prime })||\partial _{j}x^{\prime
}{}^{j}||-\mathcal{L}(x)\right] \,d^{4}x\text{.}
\end{equation}%
Since the Jacobian for the infinitesimal variation of coordinates becomes 
\begin{equation}
||\partial _{j}x^{\prime }{}^{j}||=1+\partial _{j}(\delta x^{j})\text{,}
\end{equation}%
the variation of the action takes the form, 
\begin{equation}
\delta I=\int_{_{\Omega }}\left[ \delta \mathcal{L}(x)+\mathcal{L}%
(x)\,\partial _{j}(\delta x^{j})\right] \,d^{4}x  \label{Action1}
\end{equation}%
where%
\begin{equation}
\delta \mathcal{L}(x)=\mathcal{L}^{\prime }(x^{\prime })-\mathcal{L}(x)\text{%
.}
\end{equation}

For any function $\Phi (x)$ of $x$, it is convenient to define the fixed
point variation $\delta _{0}$ by, 
\begin{equation}
\delta _{0}\Phi (x):=\Phi ^{\prime }(x)-\Phi (x)=\Phi ^{\prime }(x^{\prime
})-\Phi (x^{\prime })\text{.}
\end{equation}%
Expanding the function to first order in $\delta x^{j}$ as%
\begin{equation}
\Phi (x^{\prime })=\Phi (x)+\delta x^{j}\,\partial _{j}\Phi (x)\text{,}
\end{equation}%
we obtain 
\begin{equation}
\delta \Phi (x)=\Phi ^{\prime }(x^{\prime })-\Phi (x)=\Phi ^{\prime
}(x^{\prime })-\Phi (x^{\prime })+\Phi (x^{\prime })-\Phi (x)=\delta
_{0}\Phi (x)+\delta x^{j}\,\partial _{j}\Phi (x)\text{,}
\end{equation}%
or equivalently, 
\begin{equation}
\delta _{0}\Phi (x)=\delta \Phi (x)-\delta x^{j}\partial _{j}\Phi (x)\text{.}
\end{equation}%
An advantage of having the fixed point variation is that $\delta _{0}$
commutes with $\partial _{j}$: 
\begin{equation}
\delta _{0}\partial _{j}\Phi (x)=\partial _{j}\delta _{0}\Phi (x)\text{.}
\end{equation}%
For $\Phi (x)=\chi (x)$, we have 
\begin{equation}
\delta \chi (x)=\delta _{0}\chi (x)+\delta x^{i}\partial _{i}\chi (x)\text{,}
\end{equation}%
and%
\begin{equation}
\delta \partial _{i}\chi (x)=\partial _{i}(\delta _{0}\chi (x))-\partial
_{j}(\delta x^{j})\partial _{i}\chi (x)\text{.}
\end{equation}%
Use of the fixed point variation in the integrand of (\ref{Action1}) gives 
\begin{equation}
\delta I=\int_{_{\Omega }}\left[ \delta _{0}\mathcal{L}(x)+\partial
_{j}(\delta x^{j}\,\mathcal{L}(x))\right] \,d^{4}x\text{.}  \label{Action2}
\end{equation}%
If we require the action integral defined over any arbitrary region $\Omega $
to be invariant, that is, $\delta I=0$, then we must have 
\begin{equation}
\delta \mathcal{L}+\mathcal{L}\partial _{j}(\delta x^{j})=\delta _{0}%
\mathcal{L}+\partial _{j}(\mathcal{L}\delta x^{j})=0\text{.}
\end{equation}%
If $\partial _{j}(\delta x^{j})=0$, then $\delta \mathcal{L}=0$ so that the
Lagrangian density $\mathcal{L}$ is invariant. In general however, $\partial
_{j}(\delta x^{j})\neq 0$, and $\mathcal{L}$ transforms like a scalar
density. In other words, $\mathcal{L}$ is a Lagrangian density unless $%
\partial _{j}(\delta x^{j})=0$.

For convenience, let us introduce a function $h(x)$ that behaves like a
scalar density, namely 
\begin{equation}
\delta h(x)+h(x)\partial _{j}(\delta x^{j})=0\text{.}
\end{equation}%
We further let $\mathcal{L}(\chi (x)$, $\partial _{j}\chi (x)$; $%
x)=h(x)L(\chi (x)$, $\partial _{j}\chi (x)$; $x)$ where the function $L(\chi
(x)$, $\partial _{j}\chi (x)$; $x)$ is the scalar Lagrangian of the system.
Then we see that 
\begin{equation}
\delta \mathcal{L}+\mathcal{L}\partial _{j}(\delta x^{j})=h(x)\delta L\text{.%
}
\end{equation}%
Hence, the action integral remains invariant provided 
\begin{equation}
\delta L=0\text{.}
\end{equation}

Let us calculate the integrand of (\ref{Action2}) explicitly. The fixed
point variation of $\mathcal{L}(x)$ is a consequence of a fixed point
variation of the field $\chi (x)$, 
\begin{equation}
\delta _{0}\mathcal{L}=\frac{\partial \mathcal{L}}{\partial \chi (x)}\delta
_{0}\chi (x)+\frac{\partial \mathcal{L}}{\partial (\partial _{j}\chi (x))}%
\delta _{0}(\partial _{j}\chi (x))
\end{equation}%
which can be cast into the form, 
\begin{equation}
\delta _{0}\mathcal{L}=[\mathcal{L}]_{\chi }\delta _{0}\chi (x)+\partial
_{j}\left( \frac{\partial \mathcal{L}}{\partial (\partial _{j}\chi (x))}%
\delta _{0}\chi (x)\right)
\end{equation}%
where%
\begin{equation}
\lbrack \mathcal{L}]_{\chi }\equiv \frac{\partial \mathcal{L}}{\partial \chi
(x)}-\partial _{j}\left( \frac{\partial \mathcal{L}}{\partial (\partial
_{j}\chi (x))}\right) \text{.}
\end{equation}%
Consequently, we have the action integral in the form 
\begin{equation}
\delta I=\int_{_{\Omega }}\left\{ [\mathcal{L}]_{\chi }\delta _{0}\chi
(x)+\partial _{j}\left( \frac{\partial \mathcal{L}}{\partial (\partial
_{j}\chi (x))}\delta \chi (x)-T_{k}^{j}\,\delta x^{k}\right) \right\} d^{4}x%
\text{,}
\end{equation}%
where 
\begin{equation}
T_{k}^{j}:=\frac{\partial \mathcal{L}}{\partial (\partial _{j}\chi (x))}%
\partial _{k}\chi (x)-\delta _{k}^{j}\,\mathcal{L}  \label{energy-mom}
\end{equation}%
is the canonical energy-momentum tensor density. If the variations are
chosen in such a way that $\delta x^{j}=0$ over $\Omega $ and $\delta
_{0}\chi =0$ on the boundary of $\Omega $, then $\delta I=0$ gives us the
Euler-Lagrange equation, 
\begin{equation}
\lbrack \mathcal{L}]_{\chi }=\frac{\partial \mathcal{L}}{\partial \chi (x)}%
-\partial _{j}\left( \frac{\partial \mathcal{L}}{\partial (\partial _{j}\chi
(x))}\right) =0\text{.}
\end{equation}%
On the other hand, if the field variables obey the Euler-Lagrange equation, $%
[\mathcal{L}]_{\chi }=0$, then we have 
\begin{equation}
\partial _{j}\left( \frac{\partial \mathcal{L}}{\partial (\partial _{j}\chi
(x))}\delta \chi (x)-T^{j}\,_{k}\,\delta x^{k}\right) =0\text{,}
\end{equation}%
which gives rise to conservation laws.

\section{Global Poincar\'{e} Invariance}

Recall our assertion that our spacetime in the absence of gravitation is
Minkowski spacetime $M_{4}$. The isometry group of $M_{4}$ is the group of
Poincar\'{e} transformation (PT) which consists of the Lorentz group $SO(3$, 
$1)$ and the group of translations $T(3$, $1)$. The Poincar\'{e}
transformation of coordinates is given by 
\begin{equation}
x^{i}\overset{PT}{\rightarrow }x^{\prime }{}^{i}=a^{i}\,_{j}x^{j}+b^{i}\text{%
,}  \label{PoincareTransf}
\end{equation}%
where $a_{j}^{i}$ and $b^{i}$ are real constants and $a_{j}^{i}$ satisfy the
orthogonality conditions $a_{k}^{i}a_{j}^{k}=\delta _{j}^{i}$. For
infinitesimal variations, 
\begin{equation}
\delta x^{\prime }{}^{i}=\varepsilon ^{i}\,_{j}x^{j}+\varepsilon ^{i}
\label{PoincareTrasf1}
\end{equation}%
where $\varepsilon _{ij}+\varepsilon _{ji}=0$. While the Lorentz
transformation (LT) forms a six parameter group, the Poincar\'{e} group has
ten parameters. The Lie algebra for the ten generators of the Poincar\'{e}
group is given by%
\begin{eqnarray}
\lbrack \Xi _{ij}\text{, }\Xi _{kl}] &=&\eta _{ik}\,\Xi _{jl}+\eta
_{jl}\,\Xi _{ik}-\eta _{jk}\,\Xi _{il}-\eta _{il}\,\Xi _{jk}\text{,}  \notag
\label{Lie} \\
&&  \label{LieAlg} \\
\lbrack \Xi _{ij}\text{, }T_{k}] &=&\eta _{jk}T_{i}-\eta _{ik}T_{j}\text{, \ 
}[T_{i}\text{, }T_{j}]=0\text{,}  \notag
\end{eqnarray}%
where $\Xi _{ij}$ are the generators of Lorentz transformations, and $T_{i}$
are the generators of four-dimensional translations. Obviously, $\partial
_{i}(\delta x^{i})=0$ for the Poincar\'{e} transformation (\ref%
{PoincareTransf}). Therefore, our Lagrangian density $\mathcal{L}$, which is
the same as $L$ with $h(x)=1$ in this case, is invariant; that is to say, $%
\delta \mathcal{L}=\delta L=0$ for $\delta I=0$.

Suppose that the field $\chi (x)$ transforms under infinitesimal Poincar\'{e}
transformation as 
\begin{equation}
\delta \chi (x)=\frac{1}{2}\varepsilon ^{ij}S_{ij}\chi (x)\text{,}
\label{32}
\end{equation}%
where the tensors $S_{ij}$ are generators of the Lorentz group in some
appropriate representation, satisfying 
\begin{equation}
S_{ij}=-S_{ji}\text{, \ }[S_{ij}\text{, }S_{kl}]=\eta _{ik}\,S_{jl}+\eta
_{jl}\,S_{ik}-\eta _{jk}\,S_{il}-\eta _{il}\,S_{jk}\text{.}
\label{LorentzGen}
\end{equation}%
Correspondingly, the derivative of $\chi (x)$ transforms as 
\begin{equation}
\delta (\partial _{k}\chi (x))=\frac{1}{2}\varepsilon ^{ij}S_{ij}\partial
_{k}\chi (x)-\varepsilon ^{i}\,_{k}\partial _{i}\chi (x)\text{.}
\end{equation}%
Since the choice of infinitesimal parameters $\varepsilon ^{i}$ and $%
\varepsilon ^{ij}$ is arbitrary, the vanishing variation of the Lagrangian
density invariant $\delta \mathcal{L}=0$ leads to the following identities, 
\begin{equation}
\frac{\partial \mathcal{L}}{\partial \chi (x)}S_{ij}\chi (x)+\frac{\partial 
\mathcal{L}}{\partial (\partial _{k}\chi (x))}(S_{ij}\partial _{k}\chi
(x)+\eta _{ki}\partial _{j}\chi (x)-\eta _{kj}\partial _{i}\chi (x))=0\text{.%
}
\end{equation}%
We also obtain the following conservation laws 
\begin{equation}
\partial _{j}T_{k}^{j}=0\text{,}\ \partial _{k}\left(
S^{k}\,_{ij}-x_{i}T^{k}\,_{j}+x_{j}T^{k}\,_{i}\right) =0\text{,}
\end{equation}%
where 
\begin{equation}
S^{k}\,_{ij}:=-\frac{\partial \mathcal{L}}{\partial (\partial _{k}\chi (x))}%
S_{ij}\chi (x)
\end{equation}%
and $T_{k}^{j}$ was defined in (\ref{energy-mom}). These conservation laws
imply that the energy-momentum $P_{k}$ and angular momentum $J_{ij}$,
respectively%
\begin{equation}
P_{k}=\int T_{k}^{0}\,d^{3}x\text{,}\ J_{ij}=\int \left[ S^{0}\,_{ij}\,-%
\left( x_{i}T^{0}\,_{j}-x_{j}T^{0}\,_{i}\right) \right] d^{3}x\text{,}
\end{equation}%
are conserved. The system exhibiting invariance under the ten parameter
symmetry group has ten conserved quantities. This is an example of Noether's
theorem. The first term $S^{0}\,_{ij}\,$of the angular momentum integral
corresponds to the spin angular momentum while the second term gives the
orbital angular momentum. The global Poincar\'{e} invariance of a system
defined over spacetime implies the latter is homogeneous (all spacetime
points are equivalent) as dictated by translational invariance and is
isotropic (all directions about a spacetime point are equivalent) as
indicated by Lorentz invariance. It is interesting to observe that the fixed
point variation of the field variables $\chi (x)$ takes the form 
\begin{equation}
\delta _{0}\chi (x)=\frac{1}{2}\varepsilon ^{j}\,_{k}\Xi _{j}\,^{k}\,\chi
(x)+\varepsilon ^{j}\,T_{j}\,\chi (x)\text{,}
\end{equation}%
with $\Xi _{j}\,^{k}=\eta ^{ik}\Xi _{ji}$, where 
\begin{equation}
\Xi _{jk}=S_{jk}+\delta \left( x_{j}\partial _{k}-x_{k}\partial _{j}\right) 
\text{, \ }T_{j}=-\partial _{j}\text{.}
\end{equation}%
We remark that $\Xi _{j}\,^{k}$ are the generators of the Lorentz
transformation and $T_{j}$ are those of the translations.

\section{Local Poincar\'{e} Invariance}

In this Section we consider a modification of the infinitesimal Poincar\'{e}
transformation (\ref{PoincareTrasf1}) by assuming that the parameters $%
\varepsilon _{k}^{j}$ and $\varepsilon ^{j}$ are functions of spacetime
coordinates. We write the spacetime dependant infinitesimal Poincar\'{e}
transformation as 
\begin{equation}
\delta x^{\mu }=\varepsilon ^{\mu }\,_{\nu }(x)\,x^{\nu }+\varepsilon ^{\mu
}(x)=\xi ^{\mu }(x)\text{,}
\end{equation}%
which we call a local Poincar\'{e} transformation (or the general coordinate
transformation). To make a distinction between global (with holonomic
coordinates) and local transformations (with non-holonomic coordinates), we
use Greek indices $(\mu $, $\nu =0$, $1$, $2$, $3)$ for the former and Latin
indices $(j$, $k=0$, $1$, $2$, $3)$ for the latter. The variation of the
field variables $\chi (x)$ defined at a point $x$ is still the same as that
of the global Poincar\'{e} transformation (\ref{32}). The corresponding
fixed point variation of $\chi (x)$ takes the form, 
\begin{equation}
\delta _{0}\chi (x)=\frac{1}{2}\varepsilon _{ij}S^{ij}\chi (x)-\xi ^{\nu
}\partial _{\nu }\chi (x)\text{.}  \label{FixedPtVar}
\end{equation}%
Differentiating both sides of (\ref{FixedPtVar}) with respect to $x^{\mu }$
we obtain 
\begin{equation}
\delta _{0}\partial _{\mu }\chi (x)=\frac{1}{2}\varepsilon
^{ij}S_{ij}\partial _{\mu }\chi (x)+\frac{1}{2}(\partial _{\mu }\varepsilon
^{ij})\,S_{ij}\chi (x)-\partial _{\mu }(\xi ^{\nu }(x)\partial _{\nu }\chi
(x))\text{.}
\end{equation}%
Use of these variations leads to the variation of the Lagrangian $L$, 
\begin{equation}
\delta \mathcal{L}+\partial _{\mu }(\delta x^{\mu })\mathcal{L}=h(x)\delta
L=\delta _{0}\mathcal{L}+\partial _{\nu }(\mathcal{L}\delta x^{\nu })=-\frac{%
1}{2}(\partial _{\mu }\varepsilon ^{ij})\,S^{\mu }\,_{ij}-\left( \partial
_{\mu }\xi ^{\nu }(x)\right) T_{\text{ }\nu }^{\mu \,}
\end{equation}%
which is no longer zero unless the parameters $\varepsilon ^{ij}$ and $\xi
^{\nu }(x)$ become constants. Accordingly, the action integral for the given
Lagrangian density $\mathcal{L}$ is not invariant under local Poincar\'{e}
transformation. We notice that while $\partial _{j}(\delta x^{j})=0$ for the
local Poincar\'{e} transformation, $\partial _{\mu }\xi ^{\mu }(x)$ does not
vanish under local Poincar\'{e} transformations. Hence, as expected $%
\mathcal{L}$ is not a Lagrangian scalar but a Lagrangian density. As
mentioned earlier, for defining the Lagrangian $L$ we have to select an
appropriate non-trivial scalar function $h(x)$ satisfying 
\begin{equation}
\delta h(x)+h(x)\partial _{\mu }\xi ^{\mu }(x)=0\text{.}
\end{equation}

Now we consider a minimal modification of the Lagrangian so as to make the
action integral invariant under the local Poincar\'{e} transformation. It is
rather obvious that if there is a covariant derivative $\nabla _{k}\chi (x)$
which transforms as 
\begin{equation}
\delta (\nabla _{k}\chi (x))=\frac{1}{2}\varepsilon ^{ij}S_{ij}\nabla
_{k}\chi (x)-\varepsilon ^{i}\,_{k}\nabla _{i}\chi (x)\text{,}
\end{equation}%
then a modified Lagrangian $L^{\prime }(\chi (x)$, $\partial _{k}\chi (x)$, $%
x)=L(\chi (x)$, $\nabla _{k}\chi (x)$, $x)$, obtained by replacing $\partial
_{k}\chi (x)$ in $L(\chi (x)$, $\partial _{k}\chi (x)$, $x)$ by $\nabla
_{k}\chi (x)$, remains invariant under the local Poincar\'{e}
transformation, that is 
\begin{equation}
\delta L^{\prime }=\frac{\partial L^{\prime }}{\partial \chi (x)}\delta \chi
(x)+\frac{\partial L^{\prime }}{\partial (\nabla _{k}\chi (x))}\delta
(\nabla _{k}\chi (x))=0\text{.}
\end{equation}%
To find such a $k$-covariant derivative, we introduce the gauge fields $%
A^{ij}\,_{\mu }=-A^{ji}\,_{\mu }$ and define the $\mu $-covariant derivative 
\begin{equation}
\nabla _{\mu }\chi (x):=\partial _{\mu }\chi (x)+\frac{1}{2}A^{ij}\,_{\mu
}S_{ij}\chi (x)\text{,}  \label{mu-covariant}
\end{equation}%
in such a way that the covariant derivative transforms as 
\begin{equation}
\delta _{0}\nabla _{\mu }\chi (x)=\frac{1}{2}\varepsilon ^{ij}S_{ij}\nabla
_{\mu }\chi (x)-\partial _{\mu }(\xi ^{\nu }(x)\nabla _{\nu }\chi (x))\text{.%
}
\end{equation}%
The transformation properties of $A_{\;\;\;\mu }^{ab}$ are determined by $%
\nabla _{\mu }\chi (x)$ and $\delta \nabla _{\mu }\chi (x)$. Making use of 
\begin{equation}
\delta \nabla _{\mu }\chi (x)=\frac{1}{2}\partial _{\mu }\left( \varepsilon
^{ij}\right) S_{ij}\chi (x)+\frac{1}{2}\varepsilon ^{ij}S_{ij}\partial _{\mu
}\chi (x)-\left( \partial _{\mu }\xi ^{\nu }(x)\right) \partial _{\nu }\chi
(x)+\frac{1}{2}\delta A_{\;\;\;\mu }^{ij}S_{ij}\chi (x)+\frac{1}{4}%
A_{\;\;\;\mu }^{ij}S_{ij}\varepsilon ^{kl}S_{kl}\chi (x)
\end{equation}%
and comparing with (\ref{mu-covariant}) we obtain,%
\begin{equation}
\delta A_{\;\;\;\mu }^{ij}S_{ij}\chi (x)+\partial _{\mu }\left( \varepsilon
^{ij}\right) S_{ij}\chi (x)+\frac{1}{2}\left( A_{\;\;\;\mu }^{ij}\varepsilon
^{kl}-\varepsilon ^{ij}A_{\;\;\;\mu }^{kl}\right) S_{ij}S_{kl}\chi
(x)+\left( \partial _{\mu }\xi ^{\nu }(x)\right) A_{\;\;\;\nu
}^{ij}S_{ij}\chi (x)=0\text{.}  \label{inter}
\end{equation}%
Using the antisymmetry in $ij$ and $kl$ to rewrite the term in parenthesis
on the right hand side (RHS) of (\ref{inter}) as $\left[ S_{ij}\text{, }%
S_{kl}\right] A_{\;\;\;\mu }^{ij}\varepsilon ^{kl}\chi $, we see the
explicit appearance of the commutator $\left[ S_{ij}\text{, }S_{kl}\right] $%
. Using the expression for the commutator of Lie algebra generators%
\begin{equation}
\left[ S_{ij}\text{, }S_{kl}\right] =\frac{1}{2}c_{\;\;\;\;\ \left[ ij\right]
\left[ kl\right] }^{\left[ ef\right] }S_{ef}\text{,}
\end{equation}%
where $c_{\;\;\;\;\left[ ij\right] \left[ kl\right] }^{\left[ ef\right] }$
(the square brackets denote anti-symmetrization) is the structure constants
of the Lorentz group (deduced below), we have%
\begin{equation}
\left[ S_{ij}\text{, }S_{kl}\right] A_{\;\;\;\mu }^{ij}\varepsilon ^{kl}=%
\frac{1}{2}\left( A_{\mu }^{ic}\varepsilon _{c}^{j}-A_{\mu }^{cj}\varepsilon
_{c}^{i}\right) S_{ij}\text{.}
\end{equation}%
Substitution of this equation and consideration of the antisymmetry of $%
\varepsilon _{c}^{\;b}=-\varepsilon _{\;c}^{b}$ enables us to write 
\begin{equation}
\delta A^{ij}\,_{\mu }=\varepsilon ^{i}\,_{k}A^{kj}\,_{\mu }+\varepsilon
^{j}\,_{k}A^{ik}\,_{\mu }-(\partial _{\mu }\xi ^{\nu }(x))A^{ij}\,_{\nu
}-\partial _{\mu }\varepsilon ^{ij}\text{.}
\end{equation}%
We require the $k$-derivative and $\mu $-derivative of $\chi (x)$ to be
linearly related as 
\begin{equation}
\nabla _{k}\chi (x)=e_{k}\,^{\mu }(x)\nabla _{\mu }\chi (x)\text{,}
\label{inter2}
\end{equation}%
where the coefficients $e_{k}\,^{\mu }(x)$ are position-dependent and behave
like a new set of field variables. From (\ref{inter2}) it is evident that $%
\nabla _{k}\chi (x)$ varies as%
\begin{equation}
\delta \nabla _{k}\chi (x)=\delta e_{k}^{\mu }\nabla _{\mu }\chi
(x)+e_{k}^{\mu }\delta \nabla _{\mu }\chi (x)\text{.}  \label{test}
\end{equation}%
Comparing (\ref{test}) with $\delta \nabla _{k}\chi (x)=\frac{1}{2}%
\varepsilon ^{ab}S_{ab}\nabla _{k}\chi (x)-\varepsilon _{\text{ }%
k}^{j}\nabla _{j}\chi (x)$ we obtain,%
\begin{equation}
e_{\alpha }^{k}\delta e_{k}^{\mu }\nabla _{\mu }\chi (x)-\left( \partial
_{\alpha }\xi ^{\nu }(x)\right) \nabla _{\nu }\chi (x)+e_{\alpha
}^{k}\varepsilon _{\text{ }k}^{j}\nabla _{j}\chi (x)=0\text{.}
\end{equation}%
Exploiting $\delta \left( e_{\alpha }^{k}e_{k}^{\mu }\right) =0$ we find the
quantity $e_{k}\,^{\mu }$ transforms according to 
\begin{equation}
\delta e_{k}\,^{\mu }=e_{k}\,^{\nu }\partial _{\nu }\xi ^{\mu
}(x)-e_{i}\,^{\mu }\varepsilon ^{i}\,_{k}.
\end{equation}%
It is also important to recognize that the inverse of $\det (e_{k}\,^{\mu })$
transforms like a scalar density as $h(x)$ does. For our minimal
modification of the Lagrangian density, we utilize this available quantity
for the scalar density $h$; namely, we let 
\begin{equation}
h(x):=[\det (e_{k}\,^{\mu })]^{-1}\text{.}
\end{equation}%
By replacing the Lagrangian density $\mathcal{L}(\chi (x)$, $\partial
_{k}\chi (x)$, $x)$, which is invariant under global Poincar\'{e}
transformation, by the Lagrangian density 
\begin{equation}
\mathcal{L}(\chi (x)\text{, }\partial _{\mu }\chi (x)\text{; }x)\rightarrow
h(x)L(\chi (x)\text{, }\nabla _{k}\chi (x))\text{,}
\end{equation}%
the action integral remains invariant under the local Poincar\'{e}
transformation. We remark that in the limiting case when we consider Poincar%
\'{e} transformations which are not spacetime dependant, $e_{k}\,^{\mu
}\rightarrow \delta _{k}^{\mu }$ so that $h(x)\rightarrow 1$. Since the
local Poincar\'{e} transformation $\delta x^{\mu }=\xi ^{\mu }(x)$ is
nothing but a generalized coordinate transformation, the newly introduced
gauge fields $e_{i}^{\lambda }$ and $A^{ij}\,_{\mu }$ can be interpreted,
respectively, as the tetrad (vierbein) fields which set the local coordinate
frame and as a local affine connection with respect to the tetrad frame.

\section{Spinors and Vectors}

We may readily define tensors (and vectors) and various algebraic operations
with tensors at a given point in the spacetime manifold. Comparison of
tensors at different points however, requires use of the affine connection
via the process of parallel transport. Introduction of spinors require use
of tetrads. In analogy with the case of vectors, comparison of bilinear
forms - constructed from spinors and their conjugates - at different
spacetime points require use of the spin connection. First, consider the
case where the multiplet field $\chi \left( x\right) $ is the Dirac field $%
\psi (x)$ which behaves like a four-component spinor under LTs and
transforms as 
\begin{equation}
\psi (x)\rightarrow \psi ^{\prime }(x^{\prime })=S(\Lambda )\psi (x)\text{,}
\end{equation}%
where $S(\Lambda )$ is an irreducible unitary representation of the Lorentz
group. Since the bilinear form $v^{k}=i\bar{\psi}\gamma ^{k}\psi $ is a
vector (where $i\in 
\mathbb{C}
$), it transforms according to%
\begin{equation}
v^{j}=\Lambda _{\text{ }k}^{j}v^{k}\text{,}
\end{equation}%
where $\Lambda _{\text{ }i}^{j}$ is a LT matrix satisfying $\Lambda
_{ij}+\Lambda _{ji}=0$. Invariance of $v^{i}$ (or covariance of the Dirac
equation) under the transformation $\psi (x)\rightarrow \psi ^{\prime
}(x^{\prime })$ leads to%
\begin{equation}
S^{-1}(\Lambda )\gamma ^{i}S(\Lambda )=\Lambda _{j}^{i}\gamma ^{j}\text{,}
\label{Transf}
\end{equation}%
where the quantities $\gamma ^{i}$ are the Dirac $\gamma $-matrices
satisfying the anti-commutator relation, 
\begin{equation}
\gamma _{i}\gamma _{j}+\gamma _{j}\gamma _{i}=\eta _{ij}\mathbf{1}\text{, }i%
\text{, }j=0...3\text{.}
\end{equation}%
Furthermore, we notice that the $\gamma $-matrices satisfy the following
properties:%
\begin{equation}
\left\{ 
\begin{array}{c}
\left( \gamma _{0}\right) ^{\dag }=-\gamma _{0}\text{, }\left( \gamma
^{0}\right) ^{2}=\left( \gamma _{0}\right) ^{2}=-1\text{, }\gamma
_{0}=-\gamma ^{0}\text{ and }\gamma _{0}\gamma ^{0}=1 \\ 
\\ 
\left( \gamma _{k}\right) ^{\dag }=\gamma _{k}\text{ ,}\left( \gamma
^{k}\right) ^{2}=\left( \gamma _{k}\right) ^{2}=1\text{; }(k=1\text{, }2%
\text{, }3)\text{ and }\gamma _{k}=\gamma ^{k} \\ 
\\ 
\gamma ^{5}:=\gamma ^{0}\gamma ^{1}\gamma ^{2}\gamma ^{3}\text{, }\left(
\gamma _{5}\right) ^{\dag }=-\gamma _{5}\text{, }\left( \gamma _{5}\right)
^{2}=-1\text{ and }\gamma ^{5}=\gamma _{5}\text{.}%
\end{array}%
\right.
\end{equation}%
The dagger operator ($\dag $) implements the complex conjugation of the
transpose of the quantity appearing to its left. We assume the
transformation $S(\Lambda )$ can be put into the form $S(\Lambda
)=e^{\Lambda _{ij}\gamma ^{ij}}$. Expanding $S(\Lambda )$ about the
identity, retaining terms to first order in infinitesimals and expanding $%
\Lambda _{ij}$ to first order in $\varepsilon _{ij}$, 
\begin{equation}
\Lambda _{ij}=\delta _{ij}+\varepsilon _{ij},\varepsilon _{ij}+\varepsilon
_{ji}=0  \label{LorentzTransf1}
\end{equation}%
we get%
\begin{equation}
S(\Lambda )=1+\frac{1}{2}\varepsilon ^{ij}\gamma _{ij}\text{.}
\label{DiracTransf}
\end{equation}%
To determine the form of $\gamma _{ij}$\ we substitute (\ref{LorentzTransf1}%
) and (\ref{DiracTransf}) into (\ref{Transf}) to obtain%
\begin{equation}
\frac{1}{2}\varepsilon _{ij}\left[ \gamma ^{ij}\text{, }\gamma ^{k}\right]
=\eta ^{ki}\varepsilon _{ji}\gamma ^{j}\text{.}  \label{inter3}
\end{equation}%
Rewriting the RHS of (\ref{inter3}) using the antisymmetry of $\varepsilon
_{ij}$ as 
\begin{equation}
\eta ^{ki}\varepsilon _{ji}\gamma ^{j}=\frac{1}{2}\varepsilon _{ij}\left(
\eta ^{ki}\gamma ^{j}-\eta ^{kj}\gamma ^{i}\right) \text{,}
\end{equation}%
yields%
\begin{equation}
\left[ \gamma ^{k}\text{, }\gamma ^{ij}\right] =\eta ^{ki}\gamma ^{j}-\eta
^{kj}\gamma ^{i}\text{.}
\end{equation}%
Assuming the solution to have the form of an antisymmetric product of two
matrices, we obtain the solution%
\begin{equation}
\gamma ^{ij}:=\frac{1}{2}\left[ \gamma ^{i}\text{, }\gamma ^{j}\right] \text{%
.}
\end{equation}%
If $\chi (x)=\psi (x)$, the group generator $S_{ij}$ appearing in (\ref%
{LorentzGen}) is identified with 
\begin{equation}
S_{ij}\equiv \gamma _{ij}=\frac{1}{2}(\gamma _{i}\gamma _{j}-\gamma
_{j}\gamma _{i}).
\end{equation}%
To be explicit, the Dirac field transforms under LT as 
\begin{equation}
\delta \psi (x)=\frac{1}{2}\varepsilon ^{ij}\gamma _{ij}\psi (x)\text{.}
\end{equation}%
The Pauli conjugate of the Dirac field is denoted $\bar{\psi}(x)$ and
defined by 
\begin{equation}
\bar{\psi}(x):=i\psi ^{\dagger }(x)\,\gamma _{0}\text{.}
\end{equation}%
The conjugate field $\bar{\psi}(x)$ transforms under LTs as, 
\begin{equation}
\delta \bar{\psi}(x)=-\bar{\psi}(x)\frac{1}{2}\varepsilon ^{ij}\gamma _{ij}%
\text{.}
\end{equation}

Under local LTs, $\varepsilon _{ab}$ becomes a function of spacetime $%
\varepsilon _{ab}\rightarrow \varepsilon _{ab}(x)$. Now unlike $\partial
_{\mu }\psi (x)$, the derivative of $\psi ^{\prime }(x^{\prime })$ is no
longer homogenous due to the occurrence of the term $\gamma ^{ab}\left[
\partial _{\mu }\varepsilon _{ab}(x)\right] \psi (x)$ in $\partial _{\mu
}\psi ^{\prime }(x^{\prime })$, which is non-vanishing unless $\varepsilon
_{ab}(x)$ is constant. When going from locally flat to curved spacetime we
must generalize $\partial _{\mu }$ to the covariant derivative $D_{\mu }$ to
compensate for this extra term, allowing to gauge the group of LTs. Thus, by
use of $D_{\mu }$ we can preserve the invariance of the Lagrangian for
arbitrary local LTs at each spacetime point%
\begin{equation}
D_{\mu }\psi ^{\prime }(x^{\prime })=S(\Lambda (x))D_{\mu }\psi (x)\text{.}
\end{equation}%
To determine the explicit form of the connection belonging to $D_{\mu }$, we
study the derivative of $S(\Lambda (x))$. The transformation $S(\Lambda (x))$
is given by%
\begin{equation}
S(\Lambda (x))=1+\frac{1}{2}\varepsilon _{ab}(x)\gamma ^{ab}\text{.}
\end{equation}%
Since $\varepsilon _{ab}(x)$ is only a function of spacetime for local
Lorentz coordinates, we express this infinitesimal LT in terms of general
coordinates only by shifting all spacetime dependence of the local
coordinates into tetrad fields as 
\begin{equation}
\varepsilon _{ab}(x)=e_{a}^{\text{ \ }\lambda }(x)e_{\text{ \ }b}^{\nu
}(x)\varepsilon _{\lambda \nu }\text{.}
\end{equation}%
Substituting this expression for $\varepsilon _{ab}(x)$, we obtain%
\begin{equation}
\partial _{\mu }\varepsilon _{ab}(x)=\partial _{\mu }\left[ e_{a}^{\text{ \ }%
\lambda }(x)e_{\text{ \ }b}^{\nu }(x)\varepsilon _{\lambda \nu }\right] 
\text{.}
\end{equation}%
However, since $\varepsilon _{\lambda \nu }$ has no spacetime dependence,
this reduces to 
\begin{equation}
\partial _{\mu }\varepsilon _{ab}(x)=e_{a}^{\text{ \ }\lambda }(x)\partial
_{\mu }e_{b\lambda }(x)-e_{b}^{\text{ \ }\nu }(x)\partial _{\mu }e_{a\nu }(x)%
\text{,}  \label{inter4}
\end{equation}%
enabling us to write%
\begin{equation}
\partial _{\mu }S(\Lambda (x))=-\frac{1}{2}\gamma ^{ab}\partial _{\mu
}\varepsilon _{ab}(x)\text{.}
\end{equation}%
According to (\ref{mu-covariant}), the covariant derivative of a Dirac
spinor (and its conjugate) is given by an equation of form 
\begin{equation}
D_{\mu }\psi (x)=\partial _{\mu }\psi (x)+\frac{1}{2}\omega ^{ij}\,_{\mu
}\gamma _{ij}\psi (x)\text{ and }D_{\mu }\bar{\psi}(x)=\partial _{\mu }\bar{%
\psi}(x)-\frac{1}{2}\bar{\psi}(x)\omega ^{ij}\,_{\mu }\gamma _{ij}\text{,}
\label{del-mu-psi}
\end{equation}%
respectively, where $\omega ^{ij}\,_{\mu }$ are the components of the
spin-connection. Using the covariant derivatives of $\psi (x)$ and $\bar{\psi%
}(x)$, we can show that 
\begin{equation}
D_{\mu }v_{j}=\partial _{\mu }v_{j}-\omega ^{i}\,_{j\mu }v_{i}\text{.}
\end{equation}%
The same covariant derivative should be used for any covariant vector $v_{k}$
under LTs. Since $\nabla _{\mu }(v_{i}v^{i})=\partial _{\mu }(v_{i}v^{i})$,
the covariant derivative for a contravariant vector $v^{i}$ must be 
\begin{equation}
D_{\mu }v^{i}=\partial _{\mu }v^{i}+\omega ^{i}\,_{j\mu }v^{j}\text{.}
\end{equation}%
Since the tetrad $e_{i}\,^{\mu }$ is a covariant vector under LTs, its
covariant derivative must transform according to the same rule.

Under local LTs, the covariant derivative itself should transform as a
scalar since it does not carry a Lorentz (Latin) index. Thus $D_{\mu }v^{i}%
\overset{\text{LT}}{\rightarrow }D_{\mu }^{\prime }v^{\prime i}=\Lambda
_{j}^{i}D_{\mu }v^{j}$ where $\Lambda _{j}^{i}:=\frac{\partial x^{i}}{%
\partial x^{j}}$. Making use of the equation for\textbf{\ }$D_{\mu }v^{i}$,%
\textbf{\ }$D_{\mu }^{\prime }v^{\prime i}$ and using $\partial _{\mu }\eta
_{ab}=0$ (since the Minkowski metric is constant) to write $\Lambda _{\text{
\ }a}^{n}\eta _{nk}\partial _{\mu }\Lambda _{b}^{\text{ \ }k}=\Lambda _{%
\text{ \ }a}^{j}\partial _{\mu }\Lambda _{jb}$, we obtain the transformation
property of the spin connection 
\begin{equation}
\omega _{\text{ \ \ }\mu }^{ab}\rightarrow \omega _{\text{ \ \ }\mu
}^{\prime ab}=\Lambda _{i}^{a}\Lambda _{j}^{b}\omega _{\text{ \ \ }\mu
}^{ij}-\left( \partial _{\mu }\Lambda _{i}^{a}\right) \Lambda ^{bi}\text{.}
\label{spin-trans}
\end{equation}

Parallel transport is a unique geometric operation that is independent of
the choice of frame. We emphasize that there is only one linear connection.
It may be expressed in either holonomic or non-holonomic frames of
reference. As will be shown, these two representations of the linear
connection are related by (\ref{an-hol}). Moreover, the linear connection
(expressed in either reference frame) is not \textit{a priori} torsion free.
Indeed, it will be shown that the linear connection does contain torsion,
the latter being defined by (\ref{torsion}).

The relative rotation of a coordinate (holonomic) basis vector $e_{\alpha }$%
\textbf{\ }is given by\textbf{\ }$dx^{\alpha }\left( \partial _{\alpha
}e_{k}^{\text{ }\gamma }+\Gamma _{\alpha \beta }^{\text{ \ \ \ }\gamma
}e_{k}^{\text{ }\beta }\right) e_{\gamma }=dx^{\alpha }\left( \nabla
_{\alpha }e_{k}^{\text{ }\beta }\right) e_{\beta }^{\text{ }j}e_{j}$ with
the affine connection $\Gamma _{\;\mu \nu }^{\rho }=e_{i}^{\text{ \ }\rho
}\left( x\right) D_{\nu }e_{\text{ \ }\mu }^{i}\left( x\right) =-e_{\mu }^{%
\text{ \ }i}\left( x\right) D_{\nu }e_{\text{ \ }i}^{\rho }\left( x\right) $%
\textbf{\ }defining the covariant derivative operator $\nabla _{\alpha
}:=\partial _{\alpha }+\Gamma _{\alpha }^{\beta \gamma }\Xi _{\beta \gamma }$%
.\textbf{\ }The matrices $\Xi _{\alpha \beta }=-\Xi _{\beta \alpha }$\ are
generators of the Lorentz group satisfying the Lie algebra (\ref{LieAlg}).
To make the transition to curved spacetime, we take account of the general
coordinates of objects that are covariant under local Poincar\'{e}
transformations. The covariant derivative of a quantity $v^{\lambda }$ 
\textbf{(}$v_{\lambda }$\textbf{) }which behaves like a contravariant 
\textbf{(}covariant\textbf{) }vector under the local Poincar\'{e}
transformations is given by 
\begin{equation}
\nabla _{\nu }v^{\lambda }:=\partial _{\nu }v^{\lambda }+\Gamma ^{\lambda
}\,_{\mu \nu }v^{\mu }\text{ and }\nabla _{\nu }v_{\mu }:=\partial _{\nu
}v_{\mu }-\Gamma ^{\lambda }\,_{\mu \nu }v_{\lambda }\text{,}
\end{equation}%
respectively. The covariant derivative for a mixed tensor $A_{\nu }^{\text{
\ }\lambda }$ is given by,%
\begin{equation}
\nabla _{\mu }A_{\nu }^{\text{ \ }\lambda }=\partial _{\mu }A_{\nu }^{\text{
\ }\lambda }+\Gamma _{\text{ \ }\mu \sigma }^{\lambda }A_{\nu }^{\text{ \ }%
\sigma }-\Gamma _{\text{ \ }\mu \nu }^{\sigma }A_{\sigma }^{\text{ \ }%
\lambda }\text{.}
\end{equation}%
Since the basis vectors (in either holonomic or non-holonomic frames) change
from one point in the spacetime manifold to another, the derivative of a
vector must be given by\textbf{\ }\cite{crawford} $\partial _{\mu
}v=\partial _{\mu }\left( v^{i}e_{i}\right) =\left( \partial _{\mu
}v^{i}\right) e_{i}+v^{i}\left( \partial _{\mu }e_{i}\right) \equiv \left(
D_{\mu }v^{i}\right) e_{i}$.\ This implies that $\partial _{\mu
}e_{j}=\omega _{\text{ \ }j\mu }^{i}e_{i}$. For similar reasons, we conclude 
$\partial _{\mu }e_{\nu }=\Gamma _{\text{ \ }\nu \mu }^{\rho }e_{\rho }$.
Thus, if we perform a transformation on (\ref{spin-trans}) which leads from
a non-holonomic to a holonomic frame, then we find \cite{Hehl3, crawford} 
\begin{eqnarray}
\partial _{\nu }e_{i}\,^{\lambda }(x)-\omega ^{k}\,_{i\nu }e_{k}\,^{\lambda
}+\Gamma ^{\lambda }\,_{\mu \nu }e_{i}\,^{\mu } &:&=\mathcal{D}_{\nu
}e_{i}\,^{\lambda }(x)=0\text{,}  \label{inter6} \\
&&  \notag \\
\partial _{\nu }e^{i}\,_{\mu }(x)+\omega ^{i}\,_{k\nu }e^{k}\,_{\mu }-\Gamma
^{\lambda }\,_{\mu \nu }e^{i}\,_{\lambda } &:&=\mathcal{D}_{\nu
}e^{i}\,_{\mu }(x)=0\text{,}  \notag
\end{eqnarray}%
since $\partial _{\mu }e_{j\nu }=\partial _{\mu }\left( e_{j}\cdot e_{\nu
}\right) =\omega _{\text{ \ }j\mu }^{i}e_{i}\cdot e_{\nu }+\Gamma _{\text{ \ 
}\nu \mu }^{\rho }e_{j}\cdot e_{\rho }=\omega _{\text{ \ }j\mu }^{i}e_{i\nu
}+\Gamma _{\text{ \ }\nu \mu }^{\rho }e_{j\rho }$. The operator $\mathcal{D}%
_{\nu }$ defined in (\ref{inter6}) is introduced for later convenience. From
(\ref{inter6}) we can deduce a relation that allows to compute the affine
connection in terms of the spin connection (and tetrad) or vice-versa,
namely \cite{crawford}, 
\begin{equation}
\Gamma ^{\sigma }\,_{\mu \nu }=e_{b}^{\text{ \ }\sigma }\left( \partial
_{\mu }e_{\text{ \ }\nu }^{b}(x)-\omega _{\text{ \ \ \ \ }\mu }^{ab}e_{a\nu
}\right) \text{.}  \label{an-hol}
\end{equation}

To determine the transformation properties of (\ref{an-hol}), we consider
the LT of the quantity $e_{b}^{\text{ \ }\sigma }\partial _{\mu }e_{\text{ \ 
}\nu }^{b}(x)$ which is given by%
\begin{eqnarray}
e_{b}^{\text{ \ }\sigma }\partial _{\mu }e_{\text{ \ }\nu }^{b}(x)
&\rightarrow &\left[ e_{b}^{\text{ \ }\sigma }\partial _{\mu }e_{\text{ \ }%
\nu }^{b}(x)\right] ^{\prime }=\Lambda _{\rho }^{\sigma }e_{b}^{\text{ \ }%
\rho }\Lambda _{\mu }^{\beta }\partial _{\beta }\left( \Lambda _{\nu
}^{\lambda }e_{\text{ \ }\lambda }^{b}(x)\right)  \label{odd1} \\
&&  \notag \\
&=&\Lambda _{\lambda }^{\sigma }\Lambda _{\mu }^{\beta }\partial _{\beta
}\Lambda _{\nu }^{\lambda }+\Lambda _{\rho }^{\sigma }\Lambda _{\mu }^{\beta
}\Lambda _{\nu }^{\lambda }e_{b}^{\text{ \ }\rho }\partial _{\beta }e_{\text{
\ }\lambda }^{b}(x)\text{,}  \notag
\end{eqnarray}%
where $\Lambda _{\text{ \ }\mu }^{\alpha }:=\frac{\partial x^{\alpha }}{%
\partial x^{\mu }}$ is a holonomic transformation matrix. By use of (\ref%
{spin-trans}) and (\ref{an-hol}), we obtain the following transformation law%
\begin{equation}
\Gamma _{\text{ \ }\mu \nu }^{\lambda }\rightarrow \Gamma _{\text{ \ }\mu
\nu }^{\prime \lambda }=\Lambda _{\text{ \ }\mu }^{\alpha }\Lambda _{\text{
\ }\nu }^{\beta }\Lambda _{\gamma }^{\text{ \ }\lambda }\Gamma _{\text{ \ }%
\alpha \beta }^{\gamma }+\Lambda _{\text{ \ }\mu }^{\alpha }\Lambda _{\rho
}^{\text{ \ }\lambda }\Lambda _{\text{ \ }\alpha \nu }^{\rho }\text{,}
\end{equation}%
\textbf{\ }where\textbf{\ }$\Lambda _{\text{ \ }\alpha \nu }^{\rho }\equiv
\partial _{\alpha }\partial _{\nu }x^{\rho }$.

The linear connection may be decomposed into its symmetric and
anti-symmetric components according to $\Gamma _{\rho \mu }^{\sigma }=%
\mathring{\Gamma}_{\rho \mu }^{\sigma }+T_{\rho \mu }^{\sigma }$\textbf{,}
where\textbf{\ }$\mathring{\Gamma}_{\text{ \ }\rho \mu }^{\sigma }=\mathring{%
\Gamma}_{\text{ \ }\mu \rho }^{\sigma }$ and\textbf{\ }$T_{\rho \mu
}^{\sigma }$\textbf{\ }is the torsion tensor defined as the asymmetric part
of the affine connection,%
\begin{equation}
T_{\text{ }\beta \gamma }^{\alpha }:=\Gamma _{\text{ }\beta \gamma }^{\alpha
}-\Gamma _{\text{ }\gamma \beta }^{\alpha }\text{.}  \label{korr}
\end{equation}%
Recalling (\ref{metric}) and using (\ref{inter6}), we may derive the
so-called metricity condition $\nabla _{\lambda }g_{\mu \nu }=\mathcal{D}%
_{\lambda }g_{\mu \nu }=\mathcal{D}_{\lambda }\left( e_{\mu }^{\;i}\left(
x\right) e_{\nu }^{\;j}\left( x\right) \eta _{ij}\right) =0$. By use of the
metricity condition and the symmetry of $\mathring{\Gamma}_{\text{ \ }\mu
\nu }^{\sigma }$ in $\mu \nu $ we can write,%
\begin{equation}
\mathring{\Gamma}_{\mu \nu }^{\rho }+\mathring{\Gamma}_{\nu \mu }^{\rho
}=-e_{b}^{\rho }e_{\nu }^{c}\left[ \left( \partial _{\mu }e_{\gamma }^{\text{
}b}(x)\right) e_{\text{ }c}^{\gamma }+\left( \partial _{\mu }e_{\text{ \ }%
c}^{\gamma }(x)\right) e_{\gamma }^{\text{ }b}\right] \text{.}
\end{equation}%
We know however, that%
\begin{equation}
\partial _{\mu }\left[ e_{\text{ \ }b}^{\lambda }(x)e_{\lambda c}(x)\right]
=e_{\lambda c}(x)\partial _{\mu }e_{\text{ \ }b}^{\lambda }(x)+e_{\lambda
b}(x)\partial _{\mu }e_{\text{ \ }c}^{\lambda }(x)+e_{b}^{\text{ \ }\rho
}(x)e_{\text{ \ }c}^{\lambda }(x)\partial _{\mu }g_{\lambda \rho }\text{.}
\end{equation}%
Letting $\lambda \rightarrow \nu $ and exchanging $b$ and $c$, we obtain%
\begin{equation}
\partial _{\mu }\left[ e_{\text{ \ }b}^{\nu }(x)e_{\nu c}(x)\right] =-e_{b}^{%
\text{ \ }\lambda }(x)e_{\text{ \ }c}^{\nu }(x)\partial _{\mu }g_{\nu
\lambda }
\end{equation}%
so that,%
\begin{equation}
\mathring{\Gamma}_{\mu \lambda \nu }+\mathring{\Gamma}_{\mu \nu \lambda
}=\partial _{\mu }g_{\nu \lambda }\text{.}  \label{inter5}
\end{equation}%
By cyclic permutation of indices in (\ref{inter5}), we obtain the
Christoffel connection coefficient of a Riemannian manifold,%
\begin{equation}
\mathring{\Gamma}_{\rho \mu }^{\sigma }:=\frac{1}{2}g^{\kappa \sigma }\left(
\partial _{\kappa }g_{\rho \mu }+\partial _{\rho }g_{\mu \kappa }-\partial
_{\mu }g_{\kappa \rho }\right) \text{.}  \label{Christoffel}
\end{equation}%
For completeness we determine the transformation law of the Christoffel
connection. Making use of $\mathring{\Gamma}_{\mu \nu }^{\lambda }e_{\lambda
}=\partial _{\mu }e_{\nu }$, where%
\begin{equation}
\partial _{\mu }e_{\nu }=\Lambda _{\text{ \ }\mu }^{\alpha }\Lambda _{\text{
\ }\nu }^{\beta }\partial _{\alpha }e_{\beta }+\Lambda _{\text{ \ }\mu
}^{\alpha }\left( \partial _{\alpha }\Lambda _{\text{ \ }\nu }^{\beta
}\right) e_{\beta }\text{,}
\end{equation}%
we can show that%
\begin{equation}
\mathring{\Gamma}_{\text{ \ }\mu \nu }^{\lambda }\rightarrow \mathring{\Gamma%
}_{\text{ \ }\mu \nu }^{\prime \lambda }=\Lambda _{\text{ \ }\mu }^{\alpha
}\Lambda _{\text{ \ }\nu }^{\beta }\Lambda _{\gamma }^{\text{ \ }\lambda }%
\mathring{\Gamma}_{\text{ \ }\alpha \beta }^{\gamma }+\Lambda _{\text{ \ }%
\mu }^{\alpha }\Lambda _{\beta }^{\text{ \ }\lambda }\Lambda _{\text{ \ }%
\alpha \nu }^{\beta }\text{.}
\end{equation}

In light of the above considerations, we may regard infinitesimal local
gauge transformations as local rotations of basis vectors belonging to the
tangent space \cite{Chang, Mansouri3} of the manifold. For this reason,
given a local frame on a tangent plane to the point $x$ on the base
manifold, we can obtain all other frames on the same tangent plane by means
of local rotations of the original basis vectors. Reversing this argument,
we observe that by knowing all frames residing in the horizontal tangent
space to a point $x$ on the base manifold enables us to deduce the
corresponding gauge group of symmetry transformations.

\section{Curvature and Torsion}

The parallel transport of a vector around an infinitesimal closed path is
proportional to the curvature of the manifold and can be obtained from the
commutator $\left[ D_{\mu }\text{, }D_{\nu }\right] \psi \left( x\right) $ 
\cite{Schouten}. By direct computation we find the second order covariant
derivative%
\begin{eqnarray}
D_{\nu }D_{\mu }\psi \left( x\right) &=&\partial _{\nu }\partial _{\mu }\psi
\left( x\right) +\frac{1}{2}S_{cd}\left[ \psi \left( x\right) \partial _{\nu
}\omega _{\mu }^{\text{ \ }cd}+\omega _{\mu }^{\text{ \ }cd}\partial _{\nu
}\psi \left( x\right) \right] +\Gamma _{\text{ \ }\mu \nu }^{\rho }D_{\rho
}\psi \left( x\right) +\frac{1}{2}S_{ef}\omega _{\nu }^{\text{ \ }%
ef}\partial _{\mu }\psi \left( x\right) +  \notag  \label{pat} \\
&&  \notag \\
&&+\frac{1}{4}S_{ef}S_{cd}\omega _{\nu }^{\text{ \ }ef}\omega _{\mu }^{\text{
\ }cd}\psi \left( x\right) \text{.}  \label{patt}
\end{eqnarray}%
Using (\ref{patt}) and a similar expression with $\mu $ and $\nu $
interchanged and noting that partial derivatives commute, we find%
\begin{equation}
\left[ D_{\mu }\text{, }D_{\nu }\right] \psi \left( x\right) =\frac{1}{2}%
S_{cd}\left[ \left( \partial _{\nu }\omega _{\text{ \ \ }\mu }^{cd}-\partial
_{\mu }\omega _{\text{ \ \ }\nu }^{cd}\right) \psi \left( x\right) \right] +%
\frac{1}{4}S_{ef}S_{cd}\left[ \left( \omega _{\text{ \ \ }\nu }^{ef}\omega _{%
\text{ \ \ }\mu }^{cd}-\omega _{\text{ \ \ }\mu }^{ef}\omega _{\text{ \ \ }%
\nu }^{cd}\right) \psi \left( x\right) \right] \text{.}
\end{equation}%
Relabeling running indices we can write,%
\begin{equation}
\frac{1}{4}S_{ef}S_{cd}\left( \omega _{\text{ \ \ }\nu }^{ef}\omega _{\text{
\ \ }\mu }^{cd}-\omega _{\text{ \ \ }\mu }^{ef}\omega _{\text{ \ \ }\nu
}^{cd}\right) \psi \left( x\right) =\frac{1}{4}\left[ S_{cd}\text{, }S_{ef}%
\right] \omega _{\text{ \ }\mu }^{ef}\omega _{\text{ \ \ }\nu }^{cd}\psi
\left( x\right) \text{.}
\end{equation}%
Using $\left\{ \gamma _{a}\text{, }\gamma _{b}\right\} =2\eta _{ab}$ to
obtain%
\begin{equation}
\left\{ \gamma _{a}\text{, }\gamma _{b}\right\} \gamma _{c}\gamma _{d}=2\eta
_{ab}\gamma _{c}\gamma _{d}\text{,}
\end{equation}%
we find the commutator of $S_{ab}$ is given by%
\begin{equation}
\left[ S_{cd}\text{, }S_{ef}\right] =\frac{1}{2}\left[ \eta _{ce}\delta
_{d}^{a}\delta _{f}^{b}-\eta _{de}\delta _{c}^{a}\delta _{f}^{b}+\eta
_{cf}\delta _{e}^{a}\delta _{d}^{b}-\eta _{df}\delta _{e}^{a}\delta _{c}^{b}%
\right] S_{ab}\text{.}  \label{inter7}
\end{equation}%
It is clear that the term in brackets on the RHS of (\ref{inter7}) is
antisymmetric in $cd$ and $ef$ and is also antisymmetric under exchange of
pairs of indices $cd$ and $ef$. Since $S_{ab}$ is antisymmetric in $ab$, so
too must be the terms in brackets, so that the commutator does not vanish.
Hence, the term in brackets is totally antisymmetric under interchange of
indices $ab$, $cd$ and $ef$ and exchange of these pairs of indices. We
identify this quantity as the structure constant \cite{DeWitt} of the
Lorentz group\ 
\begin{equation}
\left[ \eta _{ce}\delta _{d}^{a}\delta _{f}^{b}-\eta _{de}\delta
_{c}^{a}\delta _{f}^{b}+\eta _{cf}\delta _{e}^{a}\delta _{d}^{b}-\eta
_{df}\delta _{e}^{a}\delta _{c}^{b}\right] ={c_{[cd][ef]}}^{[ab]}={c^{[ab]}}%
_{[cd][ef]}\text{,}
\end{equation}%
with the aid of which we can write%
\begin{equation}
\frac{1}{4}\left[ S_{cd}\text{, }S_{ef}\right] \omega _{\text{ \ \ }\mu
}^{ef}\omega _{\nu }^{cd}\psi \left( x\right) =\frac{1}{2}S_{ab}\left[
\omega _{\text{ \ }e\nu }^{a}\omega _{\text{ \ \ }\mu }^{eb}-\omega _{\text{
\ }e\nu }^{b}\omega _{\text{ \ \ }\mu }^{ae}\right] \psi \left( x\right) 
\text{.}
\end{equation}%
Combining these results, the commutator $[D_{\mu }$, $D_{\nu }]\psi \left(
x\right) $ gives 
\begin{equation}
\lbrack D_{\mu }\text{, }D_{\nu }]\psi \left( x\right) =-\frac{1}{2}%
R^{ij}\,_{\mu \nu }S_{ij}\psi \left( x\right) \text{,}
\end{equation}%
where 
\begin{equation}
R^{i}\,_{j\mu \nu }:=\partial _{\nu }\omega ^{i}\,_{j\mu }-\partial _{\mu
}^{i}\omega \,_{j\nu }+\omega ^{i}\,_{k\nu }^{k}\omega \,_{j\mu }-\omega
^{i}\,_{k\mu }^{k}\omega \,_{j\nu }\text{.}  \label{pati2}
\end{equation}%
Using the Jacobi identities for the commutator of covariant derivatives, it
follows that the curvature $R^{i}\,_{j\mu \nu }$ the Bianchi identity 
\begin{equation}
D_{\lambda }R^{i}\,_{j\mu \nu }+D_{\mu }R^{i}\,_{j\nu \lambda }+D_{\nu
}R^{i}\,_{j\lambda \mu }=0\text{.}
\end{equation}%
Permuting indices, this can be put into the cyclic form 
\begin{equation}
\varepsilon ^{\alpha \beta \rho \sigma }D_{\beta }R_{\text{ \ }\rho \sigma
}^{ij}=0\text{,}
\end{equation}%
where $\varepsilon ^{\alpha \beta \rho \sigma }$ is the Levi-Civita
alternating symbol. Furthermore, it can be shown that $R^{ij}\,_{\mu \nu
}=\eta ^{jk}R^{i}\,_{k\mu \nu }$ is antisymmetric with respect to both pairs
of indices, 
\begin{equation}
R^{ij}\,_{\mu \nu }=-R^{ji}\,_{\mu \nu }=R^{ji}\,_{\nu \mu }=-R^{ij}\,_{\nu
\mu }\text{.}
\end{equation}%
This condition is known as the first curvature tensor identity.

To determine the analogue of $[D_{\mu }$, $D_{\nu }]\psi \left( x\right) $
in local coordinates we depart from $D_{k}\psi \left( x\right) =e_{\text{ \ }%
k}^{\mu }D_{\mu }\psi \left( x\right) $. From $D_{k}\psi \left( x\right) $
we obtain,%
\begin{equation}
D_{l}D_{k}\psi \left( x\right) =e_{\text{ \ }l}^{\nu }\left[ D_{\nu }e_{%
\text{ \ }k}^{\mu }\left( x\right) \right] D_{\mu }\psi \left( x\right) +e_{%
\text{ \ }l}^{\nu }e_{\text{ \ }k}^{\mu }D_{\nu }D_{\mu }\psi \left(
x\right) \text{.}
\end{equation}%
Permuting indices and recognizing 
\begin{equation}
e_{\mu }^{\text{ \ }a}D_{\nu }e_{\text{ \ }k}^{\mu }\left( x\right) =-e_{k}^{%
\text{ \ }\mu }D_{\nu }e_{\text{ \ }\mu }^{a}\left( x\right) \text{,}
\end{equation}%
(which follows from $D_{\nu }\left( e_{\mu }^{a}e_{k}^{\mu }\right) =0$)
which leads to,%
\begin{equation}
e_{\text{ \ }l}^{\nu }\left[ D_{\nu }e_{\text{ \ }k}^{\mu }\left( x\right) %
\right] D_{\mu }\psi \left( x\right) -e_{\text{ \ }k}^{\mu }\left[ D_{\mu
}e_{\text{ \ }l}^{\nu }\left( x\right) \right] D_{\nu }\psi \left( x\right)
=\left( e_{\text{ \ }l}^{\mu }e_{\text{ \ }k}^{\nu }-e_{\text{ \ }k}^{\mu
}e_{\text{ \ }l}^{\nu }\right) \left[ D_{\nu }e_{\mu }^{\text{ \ }a}\left(
x\right) \right] D_{a}\psi \left( x\right) \text{.}
\end{equation}%
Defining%
\begin{equation}
C_{\;\;kl}^{a}:=\left( e_{\text{ \ }k}^{\mu }e_{\text{ \ }l}^{\nu }-e_{\text{
\ }l}^{\mu }e_{\text{ \ }k}^{\nu }\right) D_{\nu }e_{\mu }^{\text{ \ }%
a}\left( x\right) \text{,}  \label{pati4}
\end{equation}%
the commutator of the $k$-covariant derivatives takes the final form \cite%
{Kibble}%
\begin{equation}
\lbrack D_{k}\text{, }D_{l}]\psi \left( x\right) =-\frac{1}{2}%
R^{ij}\,_{kl}S_{ij}\psi \left( x\right) +C^{i}\,_{kl}D_{i}\psi \left(
x\right) \text{.}  \label{defalg}
\end{equation}%
The central charge $R^{ij}{}_{kl}$ and structure functions $C_{\text{ }%
jk}^{i}$ of the deformed algebra (\ref{defalg}) are given \textbf{(}in
non-holonomic coordinates\textbf{)} by the first Cartan structure equations%
\begin{equation}
R_{\text{ \ }kl}^{ij}\left( \omega \right) :=e_{k}\,^{\mu }e_{l}\,^{\nu
}R^{ij}\,_{\mu \nu }\text{, }C_{\text{ }jk}^{i}=\left( e_{\text{ \ }j}^{\mu
}e_{\text{ \ }k}^{\nu }-e_{\text{ \ }k}^{\mu }e_{\text{ \ }j}^{\nu }\right)
D_{\nu }e_{\mu }^{\text{ \ }i}\left( x\right) \text{.}  \label{struct}
\end{equation}%
With (\ref{Christoffel}) and (\ref{pati2}) in hand, the quantity $R_{\text{
\ }ijl}^{k}$ in (\ref{struct}) can be expressed in terms of its torsion-free 
$\mathring{R}_{\text{ \ }ijl}^{k}$ and torsion dependant contributions as 
\cite{Schouten}\textbf{\ }%
\begin{equation}
R_{\text{ \ }ijl}^{k}=e_{l}^{\lambda }e_{\alpha }^{k}\left( \mathring{R}_{%
\text{ }ij\lambda }^{\alpha }+2\mathring{\nabla}_{[j}T_{\text{ }i]\lambda
}^{\alpha }+2T_{\text{ }[j|\beta }^{\alpha }T_{|i]\lambda }^{\;\;\;\;\beta
}\right) \text{,}  \label{r-chris}
\end{equation}%
where\textbf{\ }$\mathring{\nabla}_{\mu }A^{\alpha }:=\partial _{\mu
}A^{\alpha }+\mathring{\Gamma}_{\mu \beta }^{\alpha }A^{\beta }$, $\mathring{%
\nabla}_{\mu }A_{\alpha }:=\partial _{\mu }A_{\alpha }-\mathring{\Gamma}%
_{\mu \alpha }^{\beta }A_{\beta }$, the square brackets in $T_{\text{ }%
[j|\beta }^{\alpha }T_{|i]\lambda }^{\;\;\;\;\beta }$ represents
anti-symmetrization with respect to $ij$, $\beta $ being fixed. As was done
for $R^{i}\,_{j\mu \nu }$ using the Jacobi identities for the commutator of
covariant derivatives, we find the Bianchi identity in Einstein-Cartan
spacetime \cite{Hehl2},%
\begin{equation}
\varepsilon ^{\alpha \beta \rho \sigma }D_{\beta }R_{\text{ }\rho \sigma
}^{\mu \nu }=\varepsilon ^{\alpha \beta \rho \sigma }C_{\beta \rho }^{\text{
\ \ \ }\lambda }R_{\text{ \ }\sigma \lambda }^{\mu \nu }\text{.}
\label{last}
\end{equation}%
It is interesting to observe the similarity in structure of the curvature
tensors in (\ref{curvature}) and the first equation in (\ref{struct}).
Indeed, there is only one curvature tensor since these two quantities can be
transformed into each other via appropriate tetrad index saturation,\textbf{%
\ }$R_{\text{ \ }jkl}^{i}\left( \omega \right) =e_{\alpha }^{i}e_{j}^{\gamma
}e_{k}^{\rho }e_{l}^{\lambda }R_{\text{ \ }\gamma \rho \lambda }^{\alpha
}\left( \Gamma \right) $. We can therefore view\textbf{\ }$R_{\text{ \ }%
\gamma \rho \lambda }^{\alpha }\left( \Gamma \right) $ in (\ref{curvature})
and\textbf{\ }$R_{\text{ \ }kl}^{ij}\left( \omega \right) $\ in\textbf{\ }(%
\ref{struct}) as holonomic and non-holonomic representations, respectively,
of the same spacetime curvature.

The second curvature identity%
\begin{equation}
R_{\text{ \ }[\rho \sigma \lambda ]}^{k}=2D_{[\rho }C_{\sigma \lambda ]}^{%
\text{ \ \ \ \ }k}-4C_{[\rho \sigma }^{\text{ \ \ \ }b}C_{\lambda ]b}^{\text{
\ \ \ }k}
\end{equation}%
leads to,%
\begin{equation}
\varepsilon ^{\alpha \beta \rho \sigma }D_{\beta }C_{\rho \sigma }^{\text{ \
\ \ }k}=\varepsilon ^{\alpha \beta \rho \sigma }R_{\text{ \ }j\rho \sigma
}^{k}e_{\text{ }\beta }^{j}\text{.}
\end{equation}%
Notice that if $\Gamma _{\;\;\mu \nu }^{\lambda }=e_{i}^{\text{ \ }\lambda
}(x)D_{\nu }e_{\text{ \ }\mu }^{i}(x)=-e_{\mu }^{\text{ \ }i}(x)D_{\nu }e_{%
\text{ \ }i}^{\lambda }(x)$, then%
\begin{equation}
\Gamma _{\;\;\mu \nu }^{\lambda }-\ \Gamma _{\;\;\nu \mu }^{\lambda
}=e_{i}^{\lambda }\left[ D_{\nu }e_{\text{ \ }\mu }^{i}\left( x\right)
-D_{\mu }e_{\text{ \ }\nu }^{i}\left( x\right) \right] \text{.}
\label{pati3}
\end{equation}%
Contracting (\ref{pati3}) with $e_{k}^{\mu }e_{l}^{\nu }$, we obtain \cite%
{Kibble}%
\begin{equation}
C_{\;\text{\ }kl}^{a}=e_{k}^{\text{ \ }\mu }e_{l}^{\text{ \ }\nu }e_{\lambda
}^{\text{ \ }a}\left( \ \Gamma _{\;\;\mu \nu }^{\lambda }-\ \Gamma _{\;\;\nu
\mu }^{\lambda }\right) \text{.}
\end{equation}%
We therefore conclude that $C_{\;kl}^{a}$ is related to the antisymmetric
part of the affine connection 
\begin{equation}
\Gamma _{\;\;\left[ \mu \nu \right] }^{\lambda }=e_{\mu }^{\text{ \ }%
k}e_{\nu }^{\text{ \ }l}e_{a}^{\text{ \ }\lambda }C_{\;\ kl}^{a}\equiv
T_{\;\;\mu \nu }^{\lambda }\text{,}  \label{torsion}
\end{equation}%
which is interpreted as spacetime torsion $T_{\;\;\mu \nu }^{\lambda }$.
Equation (\ref{torsion}) establishes a means to transform between the
holonomic torsion tensor $T_{\text{ }\beta \gamma }^{\alpha }$\ in (\ref%
{korr}) and the non-holonomic structure functions $C_{\text{ }jk}^{i}$\ in (%
\ref{struct}) (and vice-versa) in terms of appropriate tetrad index
saturation. This situation is entirely analogous to the transformation from $%
R_{\text{ \ }jkl}^{i}\left( \omega \right) $ to $R_{\text{ \ }\gamma \rho
\lambda }^{\alpha }\left( \Gamma \right) $ (and vice-versa) via tetrad index
saturation. From (\ref{pati4}), the torsion tensor can be viewed as a sort
of field strength associated with the tetrad coefficients that describes a
twist of the tetrad under parallel transport (relative to a given basis)
that is independent of the effect of curvature (i.e., a twist in a plane
perpendicular to the plane of parallel transport). This is to be compared
with the interpretation of torsion as the asymmetric part of the affine
connection. Equation (\ref{torsion}) can be solved with the aid of the
metricity condition for the spin connection, yielding \cite{Blagojevic}%
\textbf{\ }%
\begin{equation}
\omega _{ab\mu }:=\frac{1}{2}\left( \Omega _{cab}+\Omega _{bca}-\Omega
_{abc}\right) e_{\text{ \ }\mu }^{c}+T_{ab\mu }\text{,}
\end{equation}%
where 
\begin{equation}
\Omega _{cab}:=e_{\nu c}\left[ e_{\text{ \ }a}^{\mu }\partial _{\mu }e_{%
\text{ \ }b}^{\nu }(x)-e_{\text{ \ }b}^{\mu }\partial _{\mu }e_{\text{ \ }%
a}^{\nu }(x)\right] \text{,}
\end{equation}%
are the so-called objects of non-holonomicity. If the integrability
conditions $\partial _{\lbrack \alpha }e_{\beta ]}^{\text{ \ \ \ }i}=0$ are
not satisfied, the reference frame formed by $e_{i}^{\;\beta }$\ and $%
e_{\lambda }^{\;i}$ is said to be non-holonomic. The objects of
non-holonomicity measures the non-commutativity of the tetrad basis \cite%
{Hehl3}. The quantities $T_{ab\mu }$ are related to the spacetime torsion
tensor $T_{\alpha \beta \mu }$ according to $T_{ab\mu }:=e_{a}^{\text{ }%
\alpha }e_{b}^{\beta }T_{\alpha \beta \mu }$.

The most general connection in the Poincar\'{e} gauge approach to
gravitation is given by%
\begin{equation}
A_{ab\mu }=\mathring{\omega}_{ab\mu }-K_{ab\mu }+\mathring{\Gamma}^{\lambda
}\,_{\nu \mu }e_{a\lambda }e_{b}\,^{\nu }\text{,}
\end{equation}%
where $\mathring{\omega}_{ab\mu }:=\frac{1}{2}\left( \Omega _{cab}+\Omega
_{bca}-\Omega _{abc}\right) e_{\text{ \ }\mu }^{c}$ is the torsion-free spin
connection and 
\begin{equation}
K_{abc}:=-\left( T^{\lambda }\,_{\nu \mu }-T_{\nu \mu }^{\text{ \ \ }\lambda
}+T_{\mu \text{ \ }\nu }^{\text{ \ }\lambda }\right) e_{a\lambda
}e_{b}\,^{\nu }e_{c}^{\text{ \ }\mu }
\end{equation}%
is the contortion tensor. Now, the quantity $R_{\sigma \mu \nu }^{\rho
}=e_{i}\,^{\rho }R^{i}\,_{\sigma \mu \nu }$ (expressed in holonomic
coordinates)\textbf{\ }may be written as 
\begin{equation}
R^{\rho }\,_{\sigma \mu \nu }=\partial _{\nu }\Gamma _{\text{ \ }\sigma \mu
}^{\rho }-\partial _{\mu }\Gamma _{\text{ \ }\sigma \nu }^{\rho }+\Gamma
^{\rho }\,_{\lambda \nu }\Gamma ^{\lambda }\,_{\sigma \mu }-\Gamma ^{\rho
}\,_{\lambda \mu }\Gamma ^{\lambda }\,_{\sigma \nu }\text{.}
\label{curvature}
\end{equation}%
Therefore, we can regard $R^{\rho }\,_{\sigma \mu \nu }$ as the curvature
tensor with respect the affine connection $\Gamma ^{\lambda }\,_{\mu \nu }$.
We remark that $\mathring{R}_{\text{ \ }\gamma \rho \lambda }^{\alpha }$ in (%
\ref{r-chris}) is given by $\mathring{R}_{\text{ \ }\gamma \rho \lambda
}^{\alpha }=R_{\text{ \ }\gamma \rho \lambda }^{\alpha }\left( \Gamma
\rightarrow \mathring{\Gamma}\right) $. For completeness, we note that the
Ricci tensor $R_{\mu \lambda }=R_{\mu \alpha \lambda }^{\;\;\;\;\ \alpha }$
takes the form%
\begin{equation}
R_{\mu \lambda }=\mathring{R}_{\mu \lambda }+\mathring{\nabla}_{\alpha
}T_{\mu \lambda }^{\;\;\;\alpha }-\mathring{\nabla}_{\mu }T_{\alpha \lambda
}^{\;\;\;\alpha }+T_{\alpha \beta }^{\;\;\;\alpha }T_{\mu \lambda
}^{\;\;\;\beta }-T_{\mu \beta }^{\;\;\;\alpha }T_{\alpha \lambda
}^{\;\;\;\beta }\text{,}  \label{Gen-Ricci}
\end{equation}%
where the torsion-free contribution $\mathring{R}_{\mu \lambda }$ is defined
as, 
\begin{equation}
\mathring{R}_{\mu \nu }=\partial _{\gamma }\mathring{\Gamma}_{\mu \nu
}^{\gamma }-\partial _{\nu }\mathring{\Gamma}_{\mu \gamma }^{\gamma }+%
\mathring{\Gamma}_{\mu \nu }^{\gamma }\mathring{\Gamma}_{\gamma n}^{n}-%
\mathring{\Gamma}_{\mu k}^{\gamma }\mathring{\Gamma}_{\nu \gamma }^{k}\text{.%
}
\end{equation}%
It is not difficult to show that 
\begin{equation}
\sqrt{-g}=[\det e^{i}\,_{\mu }]=[\det e_{i}\,^{\mu }]^{-1}\text{,}
\end{equation}%
where $g:=\det g_{\mu \nu }$. Hence we may take $\sqrt{-g}$ for the density
function $h(x)$.

\section{Field Equations for Gravitation}

The scalar curvature $R$ is obtained from the generalized Ricci tensor (\ref%
{Gen-Ricci}) as follows,%
\begin{equation}
R=R^{\nu }\,_{\nu }=\mathring{R}+\partial _{i}K_{a}^{\text{ \ }ia}-T_{a}^{%
\text{ \ }bc}K_{bc}^{\text{ \ \ }a}
\end{equation}%
where $\mathring{R}$ denotes the usual\ Ricci scalar of general relativity.
Using this scalar curvature $R$, we choose the Lagrangian density for free
Einstein-Cartan gravity with cosmological constant 
\begin{equation}
\mathcal{L}_{G}=\frac{1}{2k_{0}}\sqrt{-g}\left( \mathring{R}+\partial
_{i}K_{a}^{\text{ \ }ia}-T_{a}^{\text{ \ }bc}K_{bc}^{\text{ \ \ }a}-2\Lambda
\right) ,
\end{equation}%
where $k_{0}=\frac{8\pi G}{c^{4}}$ is a gravitational coupling constant, and 
$\Lambda $ is the cosmological constant. Observe that the second term is a
total divergence and may be ignored. The field equation can be obtained from
the total action, 
\begin{equation}
S=\int \left\{ \mathcal{L}_{\text{field}}(\chi (x)\text{, }\partial _{\mu
}\chi (x)\text{, }e_{i}\,^{\mu }\text{, }A^{ij}\,_{\mu })+\mathcal{L}%
_{G}\right\} d^{4}x\text{,}
\end{equation}%
where the Lagrangian density for a fermion field $\psi \left( x\right) $ in
curved spacetime \cite{Wald, BirrellDavies} with torsion is given by%
\begin{equation}
\mathcal{L}_{\text{field}}=\frac{1}{2}\left[ \bar{\psi}\gamma ^{a}D_{a}\psi
\left( x\right) -\left( D_{a}\bar{\psi}\left( x\right) \right) \gamma
^{a}\psi \right] \text{.}
\end{equation}%
Modifying the connection to include spin connection and contortion
contributions, the gauge covariant derivative for a spinor and adjoint
spinor is then given by,%
\begin{equation}
D_{\mu }\psi (x)=\left\{ \partial _{\mu }+\left[ \frac{1}{4}g_{\lambda
\sigma }\left( \mathring{\omega}_{\text{ \ }\mu \rho }^{\sigma }-K_{\text{ \ 
}\rho \mu }^{\sigma }\right) \gamma ^{\lambda \rho }\right] \right\} \psi (x)%
\text{, \ }D_{\mu }\bar{\psi}(x)=\partial _{\mu }\bar{\psi}(x)-\bar{\psi}(x)%
\left[ \frac{1}{4}g_{\lambda \sigma }\left( \mathring{\omega}_{\text{ \ }\mu
\rho }^{\sigma }-K_{\text{ \ }\rho \mu }^{\sigma }\right) \gamma ^{\lambda
\rho }\right] \text{.}
\end{equation}%
The variation of the field Lagrangian reads,%
\begin{equation}
\delta \mathcal{L}_{\text{field}}=\bar{\psi}\left( \delta \gamma ^{\mu
}D_{\mu }+\gamma ^{\mu }\delta \Gamma _{\mu }\right) \psi \left( x\right) 
\text{.}
\end{equation}

The field Lagrangian defined in Einstein-Cartan spacetime can be written 
\cite{Carroll, Hehl2, Hehl3, Shapiro, Blagojevic} explicitly in terms of its
Lorentzian and contortion components as 
\begin{equation}
\mathcal{L}_{\text{field}}=\frac{1}{2}\left[ \left( \mathring{D}_{\mu }\bar{%
\psi}\left( x\right) \right) \gamma ^{\mu }\psi -\bar{\psi}\gamma ^{\mu }%
\mathring{D}_{\mu }\psi \left( x\right) \right] -\frac{\hbar c}{8}K_{\mu
\alpha \beta }\bar{\psi}\left\{ \gamma ^{\mu }\text{, }\gamma ^{\alpha \beta
}\right\} \psi \text{,}
\end{equation}%
with $\mathring{D}_{\alpha }\psi \left( x\right) :=\partial _{\alpha }\psi
\left( x\right) -\frac{1}{4}\mathring{\omega}_{\alpha ij}\gamma ^{ij}\psi
\left( x\right) $ and $\mathring{D}_{\alpha }\bar{\psi}\left( x\right)
:=\partial _{\alpha }\bar{\psi}\left( x\right) +\frac{1}{4}\bar{\psi}\left(
x\right) \mathring{\omega}_{\alpha ij}\gamma ^{ij}$. Using the following
relations%
\begin{equation}
\left\{ 
\begin{array}{c}
-\frac{1}{4}K_{\mu \alpha \beta }\bar{\psi}\left\{ \gamma ^{\mu }\text{, }%
\gamma ^{\alpha \beta }\right\} \psi =\frac{1}{4}K_{\mu \alpha \beta }\bar{%
\psi}\gamma ^{\beta \alpha }\gamma ^{\mu }\psi -\frac{1}{4}K_{\mu \alpha
\beta }\bar{\psi}\gamma ^{\mu }\gamma ^{\alpha \beta }\psi \text{,} \\ 
\\ 
\gamma ^{\mu }\gamma ^{\nu }\gamma ^{\lambda }\varepsilon _{\mu \nu \lambda
\sigma }=\left\{ \gamma ^{\mu }\text{, }\gamma ^{\nu \lambda }\right\}
\varepsilon _{\mu \nu \lambda \sigma }=3!\gamma _{\sigma }\gamma _{5}\text{,}
\\ 
\\ 
\left\{ \gamma ^{\mu }\text{, }\gamma ^{\nu \lambda }\right\} =\gamma
^{\lbrack \mu }\gamma ^{\nu }\gamma ^{\lambda ]}\text{,}%
\end{array}%
\right.
\end{equation}%
we obtain 
\begin{equation}
K_{\mu \alpha \beta }\bar{\psi}\left\{ \gamma ^{\mu }\text{, }\gamma
^{\alpha \beta }\right\} \psi =\frac{1}{2i}K_{\mu \alpha \beta }\varepsilon
^{\alpha \beta \mu \nu }\left( \bar{\psi}\gamma _{5}\gamma _{\nu }\psi
\right) \text{.}
\end{equation}%
We define the contortion axial vector%
\begin{equation}
K_{\nu }:=\frac{1}{3!}\varepsilon ^{\alpha \beta \mu \nu }K_{\alpha \beta
\mu }\text{.}
\end{equation}%
Multiplying through by the axial current $j_{\nu }^{5}=\bar{\psi}\gamma
_{5}\gamma _{\nu }\psi $ we obtain,%
\begin{equation}
\left( \bar{\psi}\gamma _{5}\gamma _{\nu }\psi \right) \varepsilon ^{\alpha
\beta \mu \nu }K_{\mu \alpha \beta }=-6ij_{\nu }^{5}K^{\nu }\text{.}
\end{equation}%
The interaction between the Dirac field and torsion has been reduced to a
coupling of the fermion axial current to a torsion axial-vector $K_{\mu }$.
\ Thus, the field Lagrangian density in curved spacetime with torsion \cite%
{Carroll} becomes%
\begin{equation}
\mathcal{L}_{\text{field}}=\frac{1}{2}\left[ \left( \mathring{D}_{\mu }\bar{%
\psi}\left( x\right) \right) \gamma ^{\mu }\psi -\bar{\psi}\gamma ^{\mu }%
\mathring{D}_{\mu }\psi \left( x\right) \right] +\frac{3i\hbar c}{8}K_{\mu
}j_{5}^{\mu }\text{.}
\end{equation}%
The total action reads,%
\begin{equation}
\delta I=\delta \int \mathcal{L}_{G}\sqrt{-g}d^{4}x+\delta \int \mathcal{L}_{%
\text{field}}\sqrt{-g}d^{4}x=\int \left( \delta \mathcal{L}_{G}+\delta 
\mathcal{L}_{\text{field}}\right) \sqrt{-g}d^{4}x\text{.}
\end{equation}

In order to obtain the explicit form of the dynamical equations for the
fermions we recall that the Dirac $\gamma $-matrices are covariantly
constant,%
\begin{equation}
\nabla _{\kappa }\gamma _{\iota }=\partial _{\kappa }\gamma _{\iota }-\Gamma
_{\iota \kappa }^{\mu }\gamma _{\mu }+\left[ \gamma _{\iota }\text{, }\hat{%
\Gamma}_{\kappa }\right] =0\text{.}
\end{equation}%
The $4\times 4$ matrices $\hat{\Gamma}_{\kappa }$\ are real matrices used to
induce similarity transformations on quantities with spinor transformation
properties \cite{Brill}, that is $\gamma _{i}\rightarrow \gamma _{i}^{\prime
}=\hat{\Gamma}^{-1}\gamma _{i}\hat{\Gamma}$. Solving for $\hat{\Gamma}%
_{\kappa }$ leads to,%
\begin{equation}
\hat{\Gamma}_{\kappa }=\frac{1}{8}\left[ \left( \partial _{\kappa }\gamma
_{\iota }\right) \gamma ^{\iota }-\Gamma _{\text{ \ }\iota \kappa }^{\mu
}\gamma _{\mu }\gamma ^{\iota }\right] \text{.}
\end{equation}%
Variation of $\hat{\Gamma}_{\kappa }$ gives $\delta \hat{\Gamma}_{\kappa }=%
\frac{1}{8}\left[ \left( \partial _{\kappa }\delta \gamma _{\iota }\right)
\gamma ^{\iota }-\left( \delta \Gamma _{\text{ \ }\iota \kappa }^{\mu
}\right) \gamma _{\mu }\gamma ^{\iota }\right] $. Since we require the
anti-commutator condition on the gamma matrices $\gamma _{\mu }\gamma _{\nu
}+\gamma _{\nu }\gamma _{\mu }=g_{\mu \nu }\mathbf{1}$ (Dirac algebra) to
hold, the variation of the metric yields,%
\begin{equation}
2\delta g^{\mu \nu }=\{\delta \gamma ^{\mu }\text{, }\gamma ^{\nu
}\}+\{\gamma ^{\mu }\text{, }\delta \gamma ^{\nu }\}\text{.}  \label{pati6}
\end{equation}%
One solution to (\ref{pati6}) is $\delta \gamma ^{\nu }=\frac{1}{2}\gamma
_{\sigma }\delta \gamma ^{\sigma \nu }$. With the aid of this result, we can
write $\left( \partial _{\kappa }\delta \gamma _{\iota }\right) \gamma
^{\iota }=\frac{1}{2}\partial _{\kappa }\left( \gamma ^{\nu }\delta g_{\nu
\iota }\right) \gamma ^{\iota }$. Finally, exploiting the anti-symmetry in $%
\gamma _{\mu \nu }$ we obtain%
\begin{equation}
\delta \hat{\Gamma}_{\kappa }=\frac{1}{8}\left[ g_{\nu \sigma }\delta \Gamma
_{\mu \kappa }^{\text{ \ \ }\sigma }-g_{\mu \sigma }\delta \Gamma _{\nu
\kappa }^{\text{ \ \ }\sigma }\right] \gamma ^{\mu \nu }\text{.}
\end{equation}%
With the above variational relations,\ it is straightforward to show that
the equation of motion obtained from variation of the action with respect to 
$\bar{\psi}(x)$ is given by \cite{Hehl2, Hehl3},%
\begin{equation}
\gamma ^{\mu }D_{\mu }\psi \left( x\right) +\frac{3}{8}T_{\mu \nu \sigma
}\gamma ^{\lbrack \mu }\gamma ^{\nu }\gamma ^{\sigma ]}\psi \left( x\right)
=0\text{.}
\end{equation}%
It is interesting to observe that this generalized, curved spacetime Dirac
equation can be recasted into a nonlinear equation of the Heisenberg-Pauli
type \cite{Hehl2},%
\begin{equation}
\gamma ^{\mu }\mathring{D}_{\mu }\psi \left( x\right) +\frac{3}{8}\left( 
\bar{\psi}\gamma ^{\mu }\gamma _{5}\psi \right) \gamma _{\mu }\gamma
_{5}\psi \left( x\right) =0\text{.}
\end{equation}

The following calculations involving the metric tensor $g_{\mu \nu }$ and
its determinant $g=\det \left( g_{\mu \nu }\right) $ are useful. Recall that 
$gg^{\mu \nu }=\frac{\partial g}{\partial g_{\mu \nu }}$ and \ $gg_{\mu \nu
}=-\frac{\partial g}{\partial g^{\mu \nu }}$. Since 
\begin{equation}
\delta \sqrt{-g}=\frac{\partial \sqrt{-g}}{\partial g}\delta g=-\frac{\delta
g}{2\sqrt{-g}}\text{,}  \label{inter11}
\end{equation}%
where $\frac{\delta g}{\delta g_{\mu \nu }}=gg^{\mu \nu }$,\ we can write%
\begin{equation}
\delta g=gg^{\mu \nu }\delta g_{\mu \nu }\text{.}  \label{inter10}
\end{equation}%
Upon substituting (\ref{inter10}) into (\ref{inter11}), we obtain $\delta 
\sqrt{-g}=-\frac{gg^{\mu \nu }\delta g_{\mu \nu }}{2\sqrt{-g}}$. However,
since $gg^{\mu \nu }\delta g_{\mu \nu }=gg_{\mu \nu }\delta g^{\mu \nu }=%
\sqrt{-g}\sqrt{-g}g_{\mu \nu }\delta g^{\mu \nu }$, we conclude $\frac{%
gg^{\mu \nu }\delta g_{\mu \nu }}{\sqrt{-g}}=\sqrt{-g}g_{\mu \nu }\delta
g^{\mu \nu }$. Hence,%
\begin{equation}
\delta \sqrt{-g}=-\frac{1}{2}\sqrt{-g}g_{\mu \nu }\delta g^{\mu \nu }\text{.}
\end{equation}%
Writing the metric in terms of the tetrads $g^{\mu \nu }=e_{\;i}^{\mu
}e^{\nu i}$, we observe $\delta \sqrt{-g}=-\frac{1}{2}\sqrt{-g}\left( \delta
e_{\;i}^{\mu }e_{\mu }^{\;i}+e_{\nu i}\delta e^{\nu i}\right) $. By using $%
\delta e^{\nu i}=\delta \left( \eta ^{ij}e_{\;j}^{\nu }\right) =\eta
^{ij}\delta e_{\;j}^{\nu }$, we are able to deduce%
\begin{equation}
\delta \sqrt{-g}=-\sqrt{-g}e_{\mu }^{\;i}\delta e_{i}^{\;\mu }.
\end{equation}%
For the variation of the Ricci tensor $R_{i\nu }=e_{i}^{\;\mu }R_{\mu \nu }$
we have $\delta \mathring{R}_{i\nu }=\delta e_{i}^{\;\mu }\mathring{R}_{\mu
\nu }+e_{i}^{\;\mu }\delta \mathring{R}_{\mu \nu }$. In an inertial frame
the Ricci tensor reduces to $\mathring{R}_{\mu \nu }=\partial _{\nu }%
\mathring{\Gamma}_{\beta \mu }^{\beta }-\partial _{\beta }\mathring{\Gamma}%
_{\nu \mu }^{\beta }$, so that 
\begin{equation}
\delta \mathring{R}_{i\nu }=\delta e_{i}^{\;\mu }\mathring{R}_{\mu \nu
}+e_{i}^{\;\mu }\left( \partial _{\nu }\delta \mathring{\Gamma}_{\beta \mu
}^{\beta }-\partial _{\beta }\delta \mathring{\Gamma}_{\nu \mu }^{\beta
}\right) \text{.}
\end{equation}%
The second term can be converted into a surface term, so it may be ignored.
Collecting our results, we have%
\begin{equation}
\left\{ 
\begin{array}{c}
\delta g^{\mu \nu }=-g^{\mu \rho }g^{\nu \sigma }\delta g_{\rho \sigma }%
\text{,} \\ 
\\ 
\delta \sqrt{-g}=-\frac{1}{2}\sqrt{-g}g_{\mu \nu }\delta g^{\mu \nu }=-\sqrt{%
-g}e_{\mu }^{\;i}\delta e_{i}^{\;\mu }\text{,} \\ 
\\ 
\delta R_{\mu \nu }=g_{\rho \mu }\left( \mathring{D}_{\lambda }\delta 
\mathring{\Gamma}_{\text{ \ \ }\nu }^{\lambda \rho }-\mathring{D}_{\nu
}\delta \mathring{\Gamma}_{\text{ \ \ }\lambda }^{\lambda \rho }\right)
+T_{\lambda \mu }^{\text{ \ \ }\rho }\delta \mathring{\Gamma}_{\text{ \ \ }%
\rho \nu }^{\lambda }\text{, \ }\delta \mathring{R}_{i\nu }=\delta
e_{i}^{\;\mu }\mathring{R}_{\mu \nu }\text{,} \\ 
\\ 
\delta R=\mathring{R}^{\mu \nu }\delta g_{\mu \nu }+g^{\mu \nu }\left( 
\mathring{D}_{\lambda }\delta \mathring{\Gamma}_{\text{ \ \ }\mu \nu
}^{\lambda }-\mathring{D}_{\nu }\delta \mathring{\Gamma}_{\text{ \ }\mu
\lambda }^{\lambda }\right) -T_{a}^{\text{ \ }bc}\delta K_{bc}^{\text{ \ \ }%
a}\text{.}%
\end{array}%
\right.
\end{equation}%
From the above results we obtain,%
\begin{equation}
\delta I_{G}=\frac{1}{16\pi }\int \left[ 
\begin{array}{c}
\left( R_{i}^{\text{ \ }\mu }-\frac{1}{2}e_{i}^{\text{ \ }\mu }R-e_{i}^{%
\text{ \ }\mu }\Lambda \right) \delta e_{\text{ }\mu }^{i}+2g^{\rho \lambda
}T_{\mu \lambda }^{\text{ \ \ }\sigma }\delta \mathring{\Gamma}_{\text{ \ }%
\rho \sigma }^{\mu }+ \\ 
\\ 
+g^{\mu \nu }\left( \mathring{D}_{\lambda }\delta \mathring{\Gamma}_{\text{
\ \ }\mu \nu }^{\lambda }-\mathring{D}_{\nu }\delta \mathring{\Gamma}_{\text{
\ }\mu \lambda }^{\lambda }\right)%
\end{array}%
\right] \sqrt{-g}d^{4}x\text{.}
\end{equation}%
The last term in the action can be converted into a surface term, so it may
be ignored. Using the four-current $v^{\mu }$ introduced earlier, the action
for the matter fields read \cite{Brill}%
\begin{eqnarray}
\delta I_{\text{field}} &=&\int \left[ \bar{\psi}\left( x\right) \delta
\gamma ^{\mu }\mathring{D}_{\mu }\psi \left( x\right) +\bar{\psi}\left(
x\right) \gamma ^{\mu }\delta \hat{\Gamma}_{\mu }\psi \left( x\right) \right]
\sqrt{-g}d^{4}x \\
&&  \notag \\
&=&\int \left\{ 
\begin{array}{c}
\left[ \frac{1}{2}g^{\mu \nu }\bar{\psi}\left( x\right) \gamma _{i}\left( 
\mathring{D}_{\nu }\psi \left( x\right) \right) +T_{\text{ \ }\rho \sigma
}^{\mu }T_{i}^{\text{ }\rho \sigma }-\delta _{i}^{\mu }T_{\lambda \rho
\sigma }T^{\lambda \rho \sigma }\right] \delta e_{\text{ }\mu }^{i}+ \\ 
\\ 
+\frac{1}{8}\left( g^{\rho \nu }v^{\mu }-g^{\rho \mu }v^{\nu }\right) \left(
g_{\mu \sigma }\delta \mathring{\Gamma}_{\text{ \ }\nu \rho }^{\sigma
}-g_{\nu \sigma }\delta \mathring{\Gamma}_{\text{ \ }\mu \rho }^{\sigma
}\right)%
\end{array}%
\right\} \sqrt{-g}d^{4}x\text{.}  \notag
\end{eqnarray}%
Removing the derivatives of variations of the metric appearing in $\delta
\Gamma _{\text{ \ }\nu \rho }^{\sigma }$ via partial integration, and
equating to zero the coefficients of $\delta g^{\mu \nu }$ and $\delta T_{%
\text{ \ }\nu \rho }^{\sigma }$\ in the variation of the action integral, we
obtain%
\begin{eqnarray}
0 &=&\frac{1}{16\pi }\left( R_{\mu \nu }-\frac{1}{2}g_{\mu \nu }R-g_{\mu \nu
}\Lambda \right) +\left( \frac{1}{2}\bar{\psi}\left( x\right) \gamma _{\nu }%
\mathring{D}_{\mu }\psi \left( x\right) -\frac{1}{4}\mathring{D}_{\mu
}v_{\nu }\right) +  \label{Field-Eqn1} \\
&&  \notag \\
&&+\mathring{D}_{\sigma }T_{\mu \nu }^{\text{ \ \ }\sigma }+T_{\mu \rho
\sigma }T_{\nu }^{\text{ }\rho \sigma }-g_{\mu \nu }T_{\lambda \rho \sigma
}T^{\lambda \rho \sigma }  \notag
\end{eqnarray}%
and%
\begin{equation}
T_{\rho \sigma \lambda }=k_{0}\tau _{\rho \sigma \lambda }\text{,}
\label{Field-Eqn2}
\end{equation}%
where $k_{0}=\frac{8\pi G}{c^{4}}$. Equation (\ref{Field-Eqn1}) has the form
of Einstein equations%
\begin{equation}
G_{\mu \nu }-g_{\mu \nu }\Lambda =k_{0}\Sigma _{\mu \nu }\text{,}
\end{equation}%
where the Einstein tensor is given by its standard definition, 
\begin{equation}
G_{\mu \nu }=R_{\mu \nu }-\frac{1}{2}g_{\mu \nu }R  \label{Einstein}
\end{equation}%
and $\Sigma _{\mu \nu }=\Theta _{\mu \nu }+\mathfrak{T}_{\mu \nu }$ is the
generalized energy-momentum tensor. We identify $\Theta _{\mu \nu }$ as the
canonical energy-momentum tensor%
\begin{equation}
\Theta _{\text{ \ }\nu }^{\mu }=\frac{\partial \mathcal{L}_{\text{field}}}{%
\partial (\mathring{D}_{\mu }\psi \left( x\right) )}\mathring{D}_{\nu }\psi
\left( x\right) -\delta _{\text{ }\nu }^{\mu }\mathcal{L}_{\text{field}}%
\text{,}
\end{equation}%
while $\mathfrak{T}_{\mu \nu }$ is the stress form of the non-Riemannian
manifold. For the case of spinor fields being considered here, the explicit
form of the energy-momentum components \cite{Schwinger} is (after
symmetrization of corresponding canonical source terms in the Einstein
equation),%
\begin{equation}
\Theta _{\mu \nu }=-\left( \bar{\psi}\left( x\right) \gamma _{(\mu }%
\mathring{D}_{\nu )}\psi \left( x\right) -\mathring{D}_{(\mu }\bar{\psi}%
\left( x\right) \gamma _{\nu )}\psi \left( x\right) \right)
\end{equation}%
and by use of the second field equation (\ref{Field-Eqn2}), we determine%
\begin{equation}
\mathfrak{T}_{\mu \nu }=\mathring{D}_{\sigma }T_{\mu \nu }^{\text{ \ \ }%
\sigma }+T_{\mu \rho \sigma }\tau _{\nu }^{\text{ }\rho \sigma }-g_{\mu \nu
}T_{\lambda \rho \sigma }\tau ^{\lambda \rho \sigma }\text{,}
\end{equation}%
where $\tau _{\mu \nu }^{\text{ \ \ }\sigma }$ is the spin angular momentum 
\cite{Hehl2, Hehl3},%
\begin{equation}
\tau _{\mu \nu }^{\text{ \ \ }\sigma }:=\frac{\partial \mathcal{L}_{\text{%
field}}}{\partial (\mathring{D}_{\sigma }\psi \left( x\right) )}\gamma _{\mu
\nu }\psi \left( x\right) \text{.}
\end{equation}%
Explicitly, the spin angular momentum reads $\tau ^{\mu \nu \sigma }=\bar{%
\psi}\gamma ^{\lbrack \mu }\gamma ^{\nu }\gamma ^{\sigma ]}\psi $.

Although the gravitational field equation is similar in form to the Einstein
field equation, it differs from the original Einstein equations because the
present curvature tensor, containing spacetime torsion is non-Riemannian. In
particular, the Einstein tensor (\ref{Einstein}) has a non-vanishing
asymmetric component and is not divergenceless, i.e., $\mathring{D}_{\nu
}G_{\mu }^{\text{ \ }\nu }\neq 0$ as can readily be verified by use of (\ref%
{last}) and (\ref{Einstein}). Assuming the Euler-Lagrange equations for the
matter fields are satisfied, we obtain the following conservation laws for
angular momentum and energy-momentum%
\begin{equation}
\left. 
\begin{array}{c}
\mathring{D}_{\nu }\mathcal{\tau }_{ij}^{\text{ \ }\nu }=e_{\text{ \ }%
i}^{\mu }e_{\text{ \ }j}^{\nu }\Sigma _{\lbrack \mu \nu ]}\text{,} \\ 
\\ 
e_{\mu }^{\text{ \ }k}\mathring{D}_{\nu }\Sigma _{\text{ \ }\kappa }^{\nu
}=2\Sigma _{\text{ \ }\kappa }^{\nu }T_{\text{ \ }\mu \nu }^{k}+\mathcal{%
\tau }_{\text{ \ }ij}^{\nu }R_{\text{ \ }\mu \nu }^{ij}\text{.}%
\end{array}%
\right.
\end{equation}

\section{Conclusion}

In the present work, we have demonstrated how all the necessary ingredients
for a theory of gravitation can be obtained from a gauge theory of local
Poincar\'{e} symmetry. Gauge fields were obtained by requiring invariance of
the Lagrangian density under local Poincar\'{e} transformations. The
resulting Einstein-Cartan theory describes a spacetime endowed with
nonvanishing curvature and torsion. The lowest order gravitational action is
one that is linear in the curvature scalar while being quadratic in torsion.
Dirac spinors were introduced as matter sources and it was found that they
couple to gravity via the torsion stress form $\mathfrak{T}_{\mu \nu }$ as
well as through the canonical energy-momentum $\Sigma _{\mu \nu }$ tensor.
The stress form contains a torsion divergence term as well as a term similar
to an external non-spinor source to gravity. The field equations obtained
from the action by means of standard variational calculus describe a
nonlinear equation of the Heisenberg-Pauli type in the matter sector, a
gravitational field equation similar in form to the Einstein equation as
well as a constraint equation relating torsion to the spin energy potential
of matter. The Bianchi identities of Einstein-Cartan theory differ from that
of general relativity since the Riemann curvature tensor characterizing the
non-Riemannian geometry does not exhibit the symmetry properties of the
latter. In the limit of vanishing torsion however, the Bianchi identities
reduce to their usual form. The conservation laws for angular momentum and
energy-momentum were obtained. From the former, it was found that the
generalized energy-momentum tensor contains a nonvanishing anti-symmetric
component proportional to the divergence of the spin angular--momentum. For
the latter, it was found that the generalized energy-momentum tensor is
divergenceless only in the limit of vanishing torsion.

\end{document}